\documentclass[prb,  twocolumn, showpacs, superscriptaddress,longbibliography]{revtex4-1}
\pdfoutput=1
\usepackage{amsmath, amsfonts, amssymb, mathrsfs}
\usepackage{graphicx, subfigure, color}
\usepackage{bm}
\usepackage{braket}
\usepackage{epstopdf}
\usepackage{dsfont}
\usepackage{ulem} % needed for \scrap-command
\usepackage{color}
\usepackage[hyperindex]{hyperref}
\hypersetup{colorlinks, citecolor=blue, filecolor=blue, linkcolor=blue, urlcolor=blue}
\renewcommand{\underline}[1]{\uline{#1}}
\newcommand{\be}{\begin{equation}}
\newcommand{\ee}{\end{equation}}
\newcommand{\bc}{\textbf{b}}
\newcommand{\A}{\mathcal{A}}
\newcommand{\nA}{\overline{\mathcal{A}}}
\newcommand{\F}{\mathcal{F}}
\newcommand{\Ss}{\mathcal{S}}
\newcommand{\BZ}{\mathcal{BZ}}
\newcommand{\C}{\mathcal{C}}

\newcommand{\matd}[4]{\begin{pmatrix} #1 & #2 \\ #3 & #4 \end{pmatrix}}

\newcommand{\vecth}[2]{\begin{pmatrix} #1 , & #2  \end{pmatrix}}

\definecolor{orange}{rgb}{1,0.5,0}

\definecolor{grey}{rgb}{.6,.6,.6}

\begin{document}

\title{Bipartite Fluctuations and Topology of Dirac and Weyl Systems}

\author{Lo\"{i}c Herviou}
\affiliation{Department of Physics, KTH Royal Institute of Technology, Stockholm, SE-106 91 Sweden}
\author{Karyn Le Hur}
\affiliation{Centre de Physique Th\'{e}orique, \'{E}cole Polytechnique, CNRS, Universit\' e Paris-Saclay, 91128 Palaiseau, France}
\author{Christophe Mora}
\affiliation{Laboratoire Pierre Aigrain, \'Ecole Normale Sup\'erieure-PSL Research University, CNRS, Universit\'e Pierre et Marie Curie-Sorbonne Universit\'es, Universit\'e Paris Diderot-Sorbonne Paris Cit\'e, 24 rue Lhomond, 75231 Paris Cedex 05, France}

\begin{abstract}
Bipartite fluctuations can provide interesting information about entanglement properties and correlations in many-body quantum systems. We address such fluctuations in relation with the topology of Dirac and Weyl quantum systems, in situations where the relevant particle number is not conserved, leading to additional volume laws scaling with the Quantum Fisher information.  Through the example of the $p+ip$ superconductor, we build a relation between charge fluctuations and the associated winding numbers of Dirac cones in the low-energy sector. Topological aspects of the Hamiltonian in the vicinity of these points induce long-range entanglement in real space. We provide a detailed analysis of such fluctuation properties, including the role of gap anisotropy,  and discuss higher-dimensional Weyl analogues.
\end{abstract}

%\pacs{03.65.Ud, 75.30.Ds, 75.10.Jm, 75.10.Dg}
\date{September 20th, 2018}
\maketitle

\section{Introduction}

Topological phases and topological quantum phase transitions have become ubiquitous in condensed matter physics. They are characterized by significant changes in the structure of the ground state, and therefore in its entanglement, while the response of local symmetry-preserving local observables are left invariant.\\
A parallel subject of study are the different gapless systems related to the gapped topological systems, whether they describe typical critical points separating between these phases, or because they are themselves topological\citep{Haldane2004, Petrescu2012}. Two-dimensional Dirac materials (graphene, $p+ip$ superconductor\citep{Volovik2003}...) and three-dimensional Weyl semi-metals are good examples of the latter. In both cases, the Hamiltonian in momentum space has a non-trivial structure close to a point-like Fermi surface (winding of the Dirac cone, or chiral charge of the Weyl node).\\
The von Neumann entanglement entropy (vNEE) (and the entanglement spectrum) are useful tools in the study of such systems\citep{Li2008, Vidal2003, Kitaev2006, Levin2006, Stephan2009, Thomale2010, Thomale2010-2, Turner2011, DeChiara2018}: they can detect phase transitions, but also measure directly specific topological properties. Experimental measurements\citep{Cardy2011, Abanin2012, Islam2015, Pichler2016, Neill2016} of such quantities remain nonetheless challenging, in particular in solid-state settings.\\
Several observables have been proposed as approximate entanglement measurements circumventing this limitation. In this paper, we study generalized bipartite charge fluctuations\citep{Klich2006,Hsu2009, Song2011, Rachel2012, Song2012-long, Nataf2012, Klich2014, Petrescu2014, Frerot2017, herviou3}. These fluctuations can be used to detect and characterize quantum phase transitions and gapless phases in one and two-dimensions\citep{Rachel2012}. In charge conserving systems, they present strong similarities with the von Neumann entanglement entropy, such as an area law for gapped ordered phases and a logarithmic growth for gapless (quasi-)ordered phases in one dimension. They can be directly measured in cold atoms experiment\citep{Bloch2008,Mazurenko2017} and in mesoscopic systems\citep{Hsu2009, Klich2009, Song2010-2, Petrescu2014}, but also through partial susceptibility measurements in actual materials \citep{Rachel2012}.
Recently, we\citep{herviou3} generalized this approach to a family of one-dimensional topological models including the celebrated Kitaev superconducting chain\citep{Kitaev2001}, and showed that the presence of a superconducting gap results in a volume type behavior of fluctuations, related to the quantum Fisher information. In addition, sub-dominant logarithmic contributions identify critical points and topological phase transitions.

The aim of this work is to further generalize the study of bipartite charge fluctuations to non-interacting gapless semi-metals in arbitrary dimensions, with a focus on the relation between topological properties of the Hamiltonian close to the gap-closing point and universal coefficient of fluctuations.\\
As in one dimension, these conformal models can be characterized by the coefficients of subleading logarithmic terms in their vNEE. Instead of being simply quantified by one number such as the central charge, the coefficients will generically be non-trivial functions of the geometric shape of the entangling surfaces. We show that the bipartite fluctuations also present similar logarithmic terms. Although no simple relations can be extracted between coefficients appearing in vNEE and bipartite charge  fluctuations, the latter can be used to characterize the topology and the nature of the Fermi surface (FS) in gapless models.  Indeed, the scaling laws will be affected by the non-analyticities of the Hamiltonian at the FS, which in turns strongly depends on the topological properties of the model when the FS is zero-dimensional, {\it i.e.} restricted to a few isolated points. Additionally, these topological markers dominate the long-range behaviour of certain well-chosen correlators. Depending on the choice of observables, standard partial susceptibilities measurements can be used to detect this behaviour.\\

This paper is organized as follows. In Section \ref{sec:Formalism}, we first introduce the general motivation of our work: bipartite fluctuations are an interesting, experimentally measurable quantity, that bring information on the entanglement structure and properties of a studied state. We then introduce the Bogoliubov formalism used to describe the generic non-interacting model we consider, and briefly introduce three models we use for illustration: the $p+ip$ superconductor, graphene and Weyl semi-metal. After detailing the different observables, we present the central idea behind our results: non-trivial topological  properties of the Hamiltonian close to gap-closing points translate into characteristic non-analyticities in momentum-space correlation functions. In turn, these non-analyticities determine the long-range behaviour of the real-space correlation function, which will therefore carry information on the different topological invariants. We then  make an explicit link between the logarithmic anomalies that may appear in their scaling laws, and the topological structure around the Fermi surface. \\
The rest of the paper is then devoted to an application of these concepts to the study and characterisation of two families of topological models. In Section \ref{sec:Dirac-basic}, we analyze the long-range correlations and bipartite fluctuations induced by the presence of isotropic Dirac cones with arbitrary winding number. We make a link with standard results on the entanglement entropy. In Section \ref{sec:Dirac-complex}, we go beyond this simple picture to study the effects of anisotropy in the Dirac cones, and the presence of multiple cones such as in graphene. We show that, even in presence of multiple gap closing points, one can identify the topology of the semi-metals. Finally, in Section \ref{sec:HighD}, we extend these results to higher-dimensions. We consider there the example of type I Weyl semi-metals, and describe  the signatures of isotropic Weyl nodes. Generalization to arbitrary dimension and number of bands is detailed in the concluding remarks.
%, before generalizing to an arbitrary dimension and number of bands.

Most of the technical difficulties are kept in Appendices for ease of reading.

\section{Analytical formalism}\label{sec:Formalism}
In this section, we begin by introducing the general properties of bipartite charge fluctuations, and their relation to entanglement entropy and mutual information. Then, we introduce the general Bogoliubov model used in the paper, followed by three concrete models: the $p+ip$ superconductor, graphene and a Weyl semi-metal. We use them as go-to examples to demonstrate that bipartite charge fluctuations can identify gapless phases involving a single or many Dirac cones or Weyl nodes. Computation of certain correlators and bipartite charge fluctuations follows. In Section \ref{subsec:FejerKernel}, we prove the relation between non-analyticities of the Hamiltonian near the Fermi surface, and the long-range behaviour of correlators and the presence of logarithmic coefficients in  bipartite charge fluctuations. Finally, in Section \ref{subsec:QFID}, we discuss the presence of a dominant volume term in bipartite charge fluctuations appearing when charge conservation is not satisfied. The volume term exhibits non-analyticities at critical points.

\subsection{Bipartite fluctuations}\label{subsec:BF}
We consider a $d-$dimensional system, noted $\Ss$. For regularization purpose, we consider a lattice of $S$ sites, which we will then take to infinity. Let $\A$ be a subregion of $\Ss$ and  $\hat{O}$ an operator that can be written as a sum of local commuting operators acting on a unit-cell:
\begin{equation}
\hat{O}_\A = \sum\limits_{\vec{r} \in \A} \hat{O}_{\vec{r}},
\end{equation}
where $\hat{O}_{\vec{r}}$ acts on the site $\vec{r}$. We define the bipartite fluctuations of $\hat{O}$ on $\mathcal{A}$, $\mathcal{F}_{\hat{O}}(\mathcal{A})$ by\citep{Song2012-long}:
\begin{eqnarray}
\mathcal{F}_{\hat{O}}(\mathcal{A}) =\Braket{ (\hat{O}_{\mathcal{A}}- \Braket{ \hat{O}_{\mathcal{A}}})^2}=\Braket{ \hat{O}_{\mathcal{A}}^2}_c\qquad \nonumber\\
= \sum\limits_{ \vec{r}, \vec{r}' \in \mathcal{A}} \Braket{ \hat{O}_{\vec{r}} \hat{O}_{\vec{r'}}} - \Braket{ \hat{O}_{\vec{r}}}\Braket{ \hat{O}_{\vec{r'}}}= \sum\limits_{ \vec{r}, \vec{r}' \in \mathcal{A}}\Braket{ \hat{O}_{\vec{r}} \hat{O}_{\vec{r'}}}_c \label{eq:formFluc}
\end{eqnarray}
where the average is  taken in the ground state.\\
As defined above, the fluctuations are always positive. They correspond to the local variance of the operator $\hat{O}$, that is to say the second order cumulants of $\hat{O}_\A$.\\
If $\hat{O}$ commutes with the Hamiltonian, the ground state can be taken to be an eigenvalue of $\hat{O}$. Then, the fluctuations verify the following set of entropy-like properties: 
\begin{itemize}
\item the fluctuations cancel for a product state (the reciprocal statement is mathematically false but empirically true as long as there is no local conservation of the charge), 
\item they are in fact symmetric ($\F_{\hat{O}}(\A) $=$ \F_{\hat{O}}(\nA)$),
\item  they admit a weak form of sub-additivity: $\F_{\hat{O}}(\A) +\F_{\hat{O}}(\mathcal{B}) \geq \F_{\hat{O}}(\mathcal{A} \cup \mathcal{B})$ where the last term identically vanishes  if $\hat{O}$ is conserved in $\A \cup \mathcal{B}$.
\end{itemize}
Finally, we note that for non-interacting Fermi gases (with conserved particle number), a universal ratio\citep{Calabrese2012} exists between the dominant coefficient of the charge fluctuations and the vNEE (and all Renyi entropies $\Ss_{\alpha}$):
\begin{equation}
\frac{\Ss_{\alpha}(\A)}{\F_{ZZ}(\A)} =\frac{\pi^2}{6} (1+\alpha^{-1}) \label{eq:RatioFermiGas}
\end{equation} 
All of these properties break down when $\hat{O}$ is not conserved, which is the focus of this paper. In particular, the fluctuations now also depend on the total volume of $\A$. For one-dimensional systems, bipartite charge fluctuations exhibit subleading logarithmic scaling terms at phase transitions~\citep{herviou3}. For Kitaev's chain \citep{Kitaev2001}, the logarithmic coefficient verifies Eq. \ref{eq:RatioFermiGas}, up to a minus sign, and thus reveals the value of the central charge. The minus sign is a result of the underlying Ising critical model~\citep{herviou3}.

As a side remark, we note that one can introduce another observable, the mutual charge fluctuations, measuring entanglement between subsystems. In analogy with mutual information, it takes the form
\begin{align}\label{mutual_info}
\mathcal{I}_{\hat{O}}(\A, \mathcal{B}) &= \F_{\hat{O}}(\A \cup \mathcal{B}) - \F_{\hat{O}}(\A) - \F_{\hat{O}}(\mathcal{B})\\ 
&= 2 \Braket{\hat{O}_\A \hat{O}_\mathcal{B}}_c, \label{eq:MF}
\end{align}
for two disjoint subregions $\A$ and $\mathcal{B}$.
The mutual fluctuations still verify $\mathcal{I}_{\hat{O}}(\A, \mathcal{B})=0$ for all product states and are extensive:
\begin{equation}
\mathcal{I}_{\hat{O}}(\A, \mathcal{B})+\mathcal{I}_{\hat{O}}(\A, \mathcal{C})=\mathcal{I}_{\hat{O}}(\A, \mathcal{B}\cup\mathcal{C}),
\end{equation}
for  disjoint regions $\A$, $\mathcal{B}$ and $\mathcal{C}$. Volume terms cancel out from Eq.~\eqref{mutual_info} highlighting the subleading contributions. The mutual fluctuations also provide a useful bound on mutual entropy
\begin{equation}
\mathcal{I}(\A, \mathcal{B})\geq \frac{1}{8} \frac{\mathcal{I}_{\hat{O}}(\A, \mathcal{B})^2}{||\hat{O}_\A||^2||\hat{O}_\mathcal{B}||^2},
\end{equation}
far from exhaustion in the models discussed in this paper. This bound nevertheless guarantees a non-trivial algebraic decay for mutual entropy in gapless phases. One finds in fact that mutual charge fluctuation and mutual entropy share similar scaling laws.

In the rest of this article, we focus primarily on bipartite charge fluctuations. Its dominant volume term  exhibits interesting singularities at phase transitions whereas the computation of $\mathcal{I}$ is generally more involved.

\subsection{Bogoliubov formalism and observables}
\subsubsection{Bogoliubov Hamiltonian}
In this paper, we consider a generic non-interacting fermionic model with two bands, describing either a normal or superconducting semi-metal, in dimensions larger than one. The one-dimensional case is studied by us in Ref.~\onlinecite{herviou3}. We illustrate our results on Dirac semi-metals in two dimensions and in Weyl semi-metals in three dimensions. Concrete examples will be introduced in the following section.\\

The generic family we study can be recast using the Bogoliubov framework as:
\begin{equation}
H=\frac{q_e}{2}  \sum\limits_{\vec{k} \in \BZ} \Psi^\dagger_{\vec{k}} \left(\vec{n}(\vec{k}).\vec{\sigma} \right) \Psi_{\vec{k}}.\label{eq:HamGen}
\end{equation}
$\vec{\sigma}$ is the vector of Pauli matrices. $\Psi_{\vec{k}}$ is a fermionic spinor:
\begin{align}
\Psi^\dagger_k &= (c^\dagger_{\vec{k}}, c_{-\vec{k}} ) \text{ for spinless superconductors,}\\
\Psi^\dagger_k &= (c^\dagger_{\vec{k}, A}, c^\dagger_{\vec{k}, B}) \text{ for normal metals.}
\end{align}
$c^{(\dagger)}$ is a fermion annihilation (creation) operator, $A$ and $B$ note two different fermionic species, such as the sublattices for graphene. $\vec{k}$ is the wave-vector and $\BZ$ the Brillouin zone.  Finally, $q_e$ is the number of inequivalent sites in each unit cell, $1$ for spinless superconductors and $2$ for the two-band normal metal models we consider. Systems with a higher number of fermions per unit-cell will be discussed in Sec.~\ref{sec:Conclusion}. The Hamiltonian~\eqref{eq:HamGen} can be diagonalized by a Bogoliubov transform (Appendix \ref{app:bogo}), with energies $\pm || \vec{n}(\vec{k}) ||$. The Green's functions in the ground state are simply given by:
\begin{align}
\Braket{0_{\eta} | \Psi_{\vec{k}} \Psi^\dagger_{\vec{k}}| 0_\eta} = \frac{1}{2}(\mathds{1} + \tilde{\textbf{n}}(\vec{k}).\vec{\sigma})\\
\text{with } \tilde{\textbf{n}}(\vec{k})=\frac{\vec{n}(\vec{k})}{||\vec{n}(\vec{k})||}
\end{align}
$\tilde{\textbf{n}}(\vec{k})$ is the spectrally-flattened Hamiltonian\citep{Fidkowski2010-2}, with the same eigenstates, and therefore correlations, as the original Hamiltonian~\eqref{eq:HamGen}.\\

\subsubsection{Models}
Dirac semi-metals are interesting examples of two-dimensional gapless metals, which include graphene\citep{Novoselov2011} but also topological systems such as $p+ip$ superconductors\citep{Read2000, Volovik2003}. Note that our results also apply for Chern and topological insulators\citep{Haldane1988, Hasan2010, Qi2011}, and for spinful superconductors such the $d$-wave superconductors. Their Fermi surface is point-like, and they are characterized by the presence of Dirac cones around which the Hamiltonian exhibits non-trivial windings.\\

Graphene consists in normal fermions hopping on a hexagonal lattice. Noting $A$ and $B$ the two triangular sublattices, its Hamiltonian for each wave-vector $\vec{k}$ in the Brillouin zone, can be written as\citep{Wallace1947, Castro2009}
\begin{align}
h_{\text{graphene}}(\vec{k})&=-t f(\vec{k}) c^\dagger_{\vec{k}, A} c_{\vec{k}, B} + h.c.\\
&\text{ with } f(\vec{k}) = 1 + 2\cos\left(\frac{k_x}{2}\right)e^{i \frac{\sqrt{3}}{2} ky}
\end{align}
where $c_{A/B}$ are spinless fermionic annihilation operators on the sublattice $A$ or $B$ and $t$ the hopping strength. The lattice spacing is set to 1. In our Bogoliubov formalism of Eq. \ref{eq:HamGen}, it translates into:
\begin{multline}
\vec{n}(\vec{k})= -(1 +2\cos\left(\frac{k_x}{2}\right) \cos \left(\frac{\sqrt{3}}{2} ky \right),\\ 2\cos\left(\frac{k_x}{2}\right) \sin \left(\frac{\sqrt{3}}{2} ky \right), 0)
\end{multline}
The system is gapless with two different Dirac cones at $\vec{K}_{\pm}=\pm(\frac{4\pi}{3}, 0)$, with opposite winding numbers. Indeed, for $\delta \vec{k} = \vec{k}-\vec{K}_\pm \ll 1$, one can expand $\vec{n}(\vec{k})$ such that
\begin{equation}
\vec{n}(\vec{k})\approx\frac{\sqrt{3}}{2} (\pm \delta k_x, - \delta k_y, 0)
\end{equation}
The gap closes linearly with a non-trivial pseudo-spin texture that gives rise to a topological invariant: the winding number. The winding number $m_\pm$ of the Dirac cone at the point $\vec{K}_\pm$ can be defined as follows: let $C_k^\pm$ be a contour circling either $\vec{K}^\pm$, we then have:
\begin{equation}
m_\pm=  \frac{-i}{2\pi} \oint\limits_{C^\pm_k} dk  \partial_k \ln (\tilde{n}_x + i\tilde{n}_y) =  \mp 1 \label{eq:windingnumber}
\end{equation}
This winding number (or the associated Berry phase) can be inferred from quantum Hall and de Shubnikov de Haas measurements, scanning probe tunneling and Klein paradox measurements\citep{Zhang2005, Novoselov2006, Braun2008, Stander2009, Berezovsky2010, Bennaceur2015}.\\

The $p+ip$ superconductor\citep{Read2000, Volovik2003} is a two-dimensional model of fermionic superconductor with unconventional superconductivity. For convenience, we limit ourselves to a regular square lattice. Its tight-binding Hamiltonian is given by:
\begin{multline}
H_{p+ip} = - \mu \sum\limits_{\vec{r}} c^\dagger_{\vec{r}} c_{\vec{r}} - t \sum\limits_{<\vec{r}, \vec{r'}>}( c^\dagger_{\vec{r}} c_{\vec{r}'} + c^\dagger_{\vec{r}'} c_{\vec{r}})\\ +\sum\limits_{\vec{r}} -i\Delta_x (c^\dagger_{\vec{r}} c^\dagger_{\vec{r} + \vec{e}_x} - h.c.) + \Delta_y (c^\dagger_{\vec{r}} c^\dagger_{\vec{r} + \vec{e}_y} + h.c.), \label{eq:pip}
\end{multline}
where $c$ are spinless fermionic annihilation operators, $<\vec{r}, \vec{r'}>$ represent the nearest-neighbor links and $\vec{e}_{x/y}$ the lattice-defining vectors. $\Delta_x$ and $\Delta_y$ are taken to be real and positive and represent mean-field, $p$-wave superconducting pairing order parameter.  The corresponding Bogoliubov Hamiltonian is characterized by:
\begin{equation}
\vec{n}(\vec{k}) = \begin{pmatrix} 2 \Delta_x \sin k_x \\ - 2 \Delta_y \sin k_y \\ -\mu -2 t \cos k_x - 2t \cos k_y 
\end{pmatrix}
\end{equation}

Several experiments have been proposed both in mesoscopic structures and cold atoms to realize such models\citep{Zhang2008, Potter2011, Buhler2014, Liu2014, Chen2015, DiBernardo2017}. Assuming non-zero $\Delta_x$, $\Delta_y$ and $t$, the system is in a phase of normal superconductivity for $|\mu|>4t$. For $0<|\mu|<4t$, it is topological, with Chern number $\pm 1$. The system is characterized by the presence of Majorana fermions at the core of vortex excitations in real space\citep{Gurarie2005, Stone2006, Nayak2008}, and of free Majorana modes at the boundaries\citep{Read2000}.\\
At the $|\mu|=4t$ critical point, the system is gapless with a single Dirac cone at one of the time-reversal symmetric points ($\vec{k}=\vec{0}$ or $\vec{k}=(\pi, \pi)$). The winding number of the cones is still given by Eq. \ref{eq:windingnumber}. The $\mu=0$ critical point separates the two topological phases, and presents two Dirac cones with identical winding numbers at $\vec{k}=(0, \pi)$ and $\vec{k}=(\pi, 0)$.\\

Weyl semi-metals\citep{Turner2013, Hosur2013, Miranksy2015, Jia2016, Burkov2016, Armitage2018} are an ubiquitous example of three-dimensional topological semi-metals arising in solid-state physics. In these materials, the gap closes only at a finite number of momenta, the Weyl points, that can have a non-trivial topological charge (chirality), though the total band structure is topologically trivial. In the rest of the paper, we will focus mainly on effective models close to the Weyl nodes, but for completeness, a possible momentum-space tight-binding Hamiltonian for each wave-vector in the Brillouin zone for a two-band Weyl semi-metal is\citep{McCormick2017}:
\begin{equation}\label{Weyl_model}
  h_{\text{Weyl}}(\vec{k})=\Psi^\dagger_{\vec{k}} \vec{n}(\vec{k}). \vec{\sigma}\Psi_ {\vec{k}}
\end{equation}
with
\begin{equation}
\vec{n}(\vec{k}) = \begin{pmatrix} -2t_x \sin k_x \\ -2t_y \sin k_y \\ -B_z -2t \cos k_x -2t \cos k_y -2t \cos k_z
  \end{pmatrix}
\end{equation}
where $\vec{\sigma}$ is the vector of Pauli matrices,  $\Psi^\dagger_{\vec{k}}=(c^\dagger_{\vec{k}, \uparrow}, c^\dagger_{\vec{k}, \downarrow})$, $t_x$ and $t_y$ are Rashba spin-orbit couplings, and $B_z$ is a Zeeman field. For non-zero $t_x$, $t_y$, and for $|B_z+4t|<2t$, the model is a semi-metal: the gap closes linearly at $\vec{k}^{0, \pm}=(0, 0, \pm \arccos \frac{B_z+4t}{-2t})$\footnote{This model actually admits two Weyl nodes for $|B_z\pm4t|<2t$ and $|B_z|<2t$.}. The two gapless points are called Weyl nodes: the low-energy Hamiltonian close to the nodes is the Weyl Hamiltonian:
\begin{equation}
h_{\text{Weyl}}=\Psi^\dagger_{\vec{k}} (v^\pm_x k_x, v^\pm_y k_y, v^\pm_z (k_z- k_z^{0, \pm})). \vec{\sigma} \Psi_ {\vec{k}}, \label{eq:Weyl2}
\end{equation}
with $v^\pm_{x/y}=-2t_{x/y}$ and $v^\pm_{z}=-2t\sin(k_z^{0, \pm})$.
These nodes carry a topological charge: the Chern number of the Hamiltonian computed on a surface enclosing only one of the nodes is equal to $\pm1$\citep{Armitage2018}. In the limiting case we study here, it reduces to $\text{sgn} (\tilde{n}_x \tilde{n}_y \tilde{n}_z)$ computed at the nodal point.\\

In the next sections, we address how the winding number of a Dirac cone (Sections \ref{sec:Dirac-basic} and \ref{sec:Dirac-complex}) or the topological charge of a Weyl point (Section \ref{sec:HighD}) relate to the bipartite charge fluctuations.

\subsubsection{Correlators and fluctuations}
The two types of two-point correlators we consider are the simplest two- and four-fermion operators. Let us define  $\Psi_{\vec{r}}$ the fermionic spinor in real space (Fourier transform of $\Psi_{\vec{k}}$), and $\hat{\textbf{n}}(\vec{r})$ the Fourier transform of the spectrally-flattened $\tilde{\textbf{n}}(\vec{k})$:
\begin{equation}
\hat{\textbf{n}}(\vec{r}) = \frac{1}{S} \sum\limits_{\vec{k} \in \BZ} e^{i\vec{k}.\vec{r}} \tilde{\textbf{n}} (\vec{k}),
\end{equation}
where $S$ is the total number of sites in the system. The first correlator we consider is simply a well-chosen Green's function.
\begin{equation}
\C^1_{\alpha}(\vec{r}-\vec{r}') = \Braket{ \Psi^\dagger_{\vec{r}} \sigma^\alpha \Psi_{\vec{r}'}} = -\hat{\textbf{n}}_{\alpha}(\vec{r}'-\vec{r}) \label{eq:defC1}
\end{equation}
The second correlator is the building block of the bipartite fluctuations. Let us first consider a non-superconducting model. A complete basis of the (non zero) local fermionic bilinears is given by:
\begin{equation}
\frac{q_e}{2}\Psi^\dagger_{\vec{r}} \Psi_{\vec{r}}, \quad \vec{S}_{\vec{r}}=\frac{q_e}{2}\Psi^\dagger_{\vec{r}} \vec{\sigma} \Psi_{\vec{r}}. 
\end{equation}
The first one, corresponding to the total number of electrons in the unit-cell, actually globally commutes with the Hamiltonian in Eq. \ref{eq:HamGen}. Let us first consider fluctuations of the other three correlators. They correspond to the different (pseudo-)spin polarization. We define:
\begin{equation}
\C^2_{\alpha \beta}(\vec{r}-\vec{r}') = \frac{q_e^2}{4}\Braket{ \Psi^\dagger_{\vec{r}} \sigma^\alpha \Psi_{\vec{r}}\Psi^\dagger_{\vec{r}'} \sigma^\beta \Psi_{\vec{r}'}}_c. \label{eq:defC2}
\end{equation}
Computing the correlator is straightforward, thanks to Wick's theorem and charge conservation. An integral form can be given:
\begin{equation}
\C^2_{\alpha \alpha}(\vec{r})=\frac{q_e}{4}\delta_{\vec{r}, 0}+ \frac{q_e}{4} \int\limits_{BZ^2} \frac{d\vec{k} d\vec{q}}{A_{BZ}^2}\mathcal{K}(\vec{k}-\vec{q}, \vec{r}) (\tilde{n}(\vec{k}). \tilde{n}(\vec{q}))_\alpha\label{eq:GenCorrSite-continuous}
\end{equation}
where $A_{BZ}$ is the area of the Brillouin zone, the kernel $\mathcal{K}(\vec{k}, \vec{r})$ is simply $e^{i \vec{k}.\vec{r}}$, $( . )_{\alpha}$ is the Minkowski scalar product with a $-1$ carried by the coordinate $\alpha$ and $\delta_{\vec{r}, 0}$ the Kronecker symbol. As a general rule, these integrals are elliptic and not analytically  computable. It is convenient to remark that:
\begin{equation}
\C^2_{\alpha \alpha}(\vec{r})=\frac{q_e}{4}\delta_{\vec{r}, 0}+ \frac{q_e}{4}||\hat{\textbf{n}}(\vec{r})||^2_\alpha\label{eq:GenCorrSite}
\end{equation}
where $||.||^2_\alpha$ is the Minkowski norm.

For the total charge component, one obtains:
\begin{align}
\C^2_{0 0}(\vec{r}-\vec{r}') &= \frac{q_e^2}{4}\Braket{ \Psi^\dagger_{\vec{r}}  \Psi_{\vec{r}}\Psi^\dagger_{\vec{r}'} \Psi_{\vec{r}'}}_c\\
&=\frac{q_e}{4}\delta_{\vec{r},\vec{r}'}-\frac{q_e}{4}||\hat{\textbf{n}}(\vec{r}-\vec{r}')||^2
\end{align}

For superconductors, the derivations are similar. In this case however, only the component $\frac{q_e}{2}\Psi_j^\dagger \sigma^z \Psi_j$ is non-vanishing in $\vec{S}_{\vec{r}}$ - the Pauli principle enforces the others two to vanish.

The knowledge of these different correlation functions gives access to bipartite charge fluctuations when summed over lattice positions
\begin{equation}\label{fluctuations_form}
\F_{\alpha \alpha}(\A)= \sum\limits_{\vec{r}, \vec{r}' \in \A} \C^2_{\alpha \alpha}(\vec{r}-\vec{r}'),
\end{equation}
corresponding to the right-hand-side of Eq.~\eqref{eq:GenCorrSite-continuous} with the kernel
\begin{equation}
\mathcal{K}(\vec{k}, \A)=\sum\limits_{\vec{r}, \vec{r}' \in \A} e^{i \vec{k}.(\vec{r}-\vec{r}')}, \label{eq:newKernel}
\end{equation}
and the contribution $q_e V_{\A}/4$ instead of the first term. The analysis of this kernel will be fundamental to determine properties of charge fluctuations. We conclude this Section with the useful formulation:
\begin{equation}
\F_{\alpha \alpha}(\A)=\frac{q_e V_{\A}}{4} + \frac{q_e}{4}\sum\limits_{\vec{r}, \vec{r}' \in \A} ||\hat{\textbf{n}}(\vec{r}-\vec{r'})||^2_\alpha\label{eq:GenFlucA}
\end{equation}
and the definition of the Heisenberg isotropic spin-spin fluctuations:
\begin{align}
\F_{\text{Hei}}(\A)&=\sum\limits_{\vec{r}, \vec{r}' \in \A} \Braket{\vec{S}_{\vec{r}}.\vec{S}_{\vec{r}'}}_c \\
&=\frac{3q_e V_{\A}}{4} + \frac{q_e}{4}\sum\limits_{\vec{r}, \vec{r}' \in \A} ||\hat{\textbf{n}}(\vec{r}-\vec{r'})||^2\label{eq:GenFlucB}\\
&=q_e V_{\A}-\F_{00}(\A),
\end{align}
where $ \F_{00}$ are the total charge fluctuations, $V_{\A}$ is the volume of the sub-region $\A$ (the area in two dimensions). Note that one can reexpress Eqs. \ref{eq:GenFlucA} and \ref{eq:GenFlucB} in terms of entangling surfaces as shown in Eq. \ref{eq:App-Entangling}: the volume sums can be rewritten in the continuum limits as integrals over the boundary of $\A$. This formulation is particularly useful to capture the subvolumic contributions to the fluctuations.

\subsection{Fej\'er kernel properties and scaling laws} \label{subsec:FejerKernel}
\subsubsection{Scaling laws and non-analyticities}
The long-range properties of all the aforementioned observables depend on the long-range behavior of the Fourier transform of the spectrally-flattened vector  $\tilde{\textbf{n}}(\vec{k})$. In this section, we study its generic dependence on the Fermi surface, and give the universal scaling laws that appear in both the correlators $\mathcal{C}^1$ and $\mathcal{C}^2$, and the bipartite charge fluctuations.\\

It is a well-known mathematical result that the scaling of the Fourier transform of a function $g$ is directly related to its non-analyticities. The best-known example is in one dimension ($d=1$). Let $\hat{g}$ be the Fourier transform of a periodic function $g$. If $g$ is $p$-differentiable with $g^{(p)}$ continuous by part, then $\tilde{g}(r) = O(r^{-(p+1)})$. By applying these results to $\tilde{n}(\vec{k})$ and the correlators $\mathcal{C}^{1/2}$\footnote{Due to Wick theorem, it is actually valid for all operators for these non-interacting systems.}, this property immediately recovers that, for a gapped system, in the absence of long-range term in the Hamiltonian, correlations decrease faster than any power-law (exponentially).
On the other hand, semi-metal gapless models exhibit non-analyticities in $\tilde{\textbf{n}}(k)$ resulting in a power-law decay of the correlation functions in real space.\\

Similar results for multi-dimensional Fourier transform are scarce. Difficulties arise from the wider variety of singularities that exist in a higher-dimensional manifold. An important mathematical notion for classifying the correlation functions is the concept of Sobolev spaces\citep{Giovanni}, that  categorize the convergence speed of $|\hat{g}(\vec{r})|^2$ (see Appendix \ref{app:Fej}). In the rest of this paper, we investigate some physical consequences of this classification. As the semi-metals we consider have a zero-dimensional Fermi surface, non-analyticities will  generally be generated by (partial) winding around the gap-closing momenta. In particular, a topological charge will lead to a characteristic set of non-analyticities.

\subsubsection{Kernel properties for the bipartite fluctuations}
The previously derived kernel $\mathcal{K}$ naturally arises in interferences problems (but interestingly also in some derivation of the entropies\citep{Wolf2006}). It is a multi-dimensional generalization of the Fej\'er kernel, and depends on the exact geometric shape of $\A$. It verifies:
\begin{equation}
\frac{1}{V_\A}\mathcal{K}(\vec{k}, \A) \rightarrow \delta^d (\vec{k}) \text{ when $\A$ covers the entire system.}
\end{equation}
$V_\A$ is the number of sites (volume) of $\A$ and $\delta^d$ is the Kronecker symbol if we consider the lattice theory, and the $d-$dimensional Dirac delta in the continuum limit. 
This relation implies that the dominant scaling term in the bipartite fluctuations will be proportional to the volume $V_\A$ of $\A$:
\begin{equation}
  \F_{\alpha \alpha}(\A)=\frac{q_e V_{\A}}{4} + \frac{q_e V_{\A}}{4} \int\limits_{BZ} \frac{d\vec{k}}{A_{BZ}} ||\tilde{\textbf{n}}(\vec{k})||^2_\alpha +o(V_{\A})
 \end{equation}
We define
\begin{equation}
i_{\alpha \alpha}=\lim\limits_{V_\A \rightarrow + \infty}  \frac{\F_{\alpha \alpha}}{V_\A}.
\end{equation}
Due to charge conservation, $i_{00}$ vanishes. We discuss the properties of $i_{\alpha \alpha}$ in the next Section.

\subsection{Dominant scaling term and Quantum Fisher information}\label{subsec:QFID}
Given the previously obtained scaling laws, we can present a systematic physical interpretation of the coefficient of the dominant volume term. Indeed, for any observable $\hat{O}$ whose fluctuations take the form of Eq. \ref{eq:formFluc}, one trivially obtains:
\begin{equation}
i_{\hat{O}} = \frac{1}{S} \sum\limits_{\vec{r}, \vec{r}' \in \mathcal{S}} \Braket{\hat{O}_{\vec{r}}\hat{O}_{\vec{r}'}}-\Braket{\hat{O}_{\vec{r}}}\Braket{\hat{O}_{\vec{r}'}},
\end{equation}
that is to say that the volume coefficient coincides with the density of fluctuations of the total system in the thermodynamic limit,  a non-universal quantity. This coefficient will be therefore non-zero if and only if $\hat{O}$ is not globally conserved in the system. 

Remarkably, for a pure state  at $T=0$ , $i_{\hat{O}}$ is actually the Quantum Fisher Information density\citep{Helstrom,Benatti2014, Toth2014} (QFID) associated to $\hat{O}$. It has been used to characterize several transitions\citep{Wang2014,  Hauke2015, Wu2016, Ye2016, Smith2016}.\\
The QFID gives a bound on the producibility of the ground state in real or momentum space\citep{Hyllus2012, Toth2012}.
Non-interacting models with two bands, the primary focus on this paper, are always $2$-producible, implying a universal bound
\begin{equation}
i_{\hat{O}} \leq \frac{q_e}{2}. \label{eq:BoundQFID}
\end{equation}
More details are given in Appendix \ref{app:QFID}.\\

\begin{figure}
\begin{center}
\includegraphics[width=0.9\linewidth]{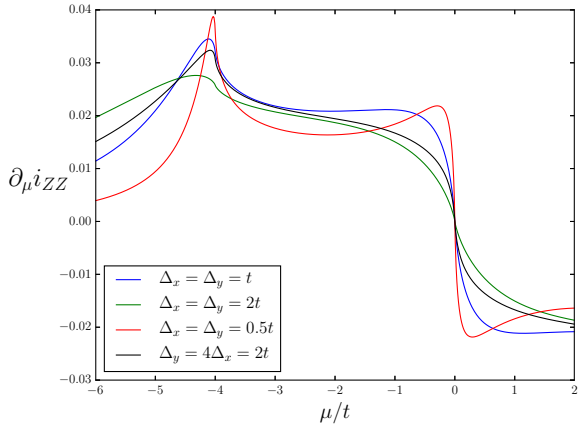}
\end{center}
\caption{Derivative of the QFID for the $p + ip$ superconductor as a function of the chemical potential for several choices of pairing order parameters $\Delta_x$ and $\Delta_y$. Phase transitions occur at $\mu=-4t$ (from trivial to topological, with one Dirac cone at $\vec{k}=0$) and at $\mu=0$ (between topological phases with opposing Chern numbers, with two Dirac cones at $\vec{k}=(0, \pi)$ and $\vec{k}=(\pi, 0)$) }
\label{fig1}
\end{figure}

$i_{\hat{O}}$ presents characteristic non-analyticities at phase transitions. As an example, $i_{ZZ}$ in a Kitaev's wire has a discontinuity of its derivative at the phase transition between the topological and the non-topological phases~\citep{herviou3}. Similarly, for the $p+ip$ superconductor introduced in Eq. \ref{eq:pip}, the second derivative of the QFID presents logarithmic divergences at the phase transitions, when varying the chemical potential, as shown in Figure \ref{fig1}. As a general rule, when crossing a phase transition while varying the chemical potential, and if the critical phase has a point-like Fermi-surface with linear dispersion, the $d^{th}$ derivative of the QFID should be discontinuous for odd physical dimensions and diverge logarithmically in even dimensions.\\

A similar increase of order has been observed in other thermodynamic quantities for these topological transitions\citep{Kempkes2016}, and is a simple consequence of the point-like nature of the Fermi surface. Note that, due to the general non-conservation of the charges we observe, the usual relation between charge fluctuations and susceptibility\citep{Bell1963} at finite temperature is no longer valid. This is both an advantage: the analytical structure of the fluctuations is much simpler, and has a more explicit dependency on the topology of the Fermi surface, and a drawback: they become more challenging to experimentally measure compared to susceptibilities or compressibilities. The total charge fluctuations $\F_{00}$ are the exception to that rule, and should therefore be more easily accessible. Fluctuations are also numerically useful, as they are straightforward to compute in most simulations scheme and allow access to the Luttinger parameter with high accuracy in Luttinger Liquids\citep{Rachel2012, Petrescu2017}.

\section{Signature of winding number for a single Dirac cone}\label{sec:Dirac-basic}
In this Section, we study a system that presents only a single Dirac cone, and generalize in the  following Section. We first present some general properties of the long-range behaviour of the chosen correlators and their relation with the Fermi surface topology. Then, we focus on charge fluctuations. The non-trivial winding of the Dirac cone leads to characteristic non-analyticities of the Hamiltonian, which in turns implies the following scaling laws for charge fluctuations:
\begin{equation}\label{scaling_flu}
\F_{\alpha \alpha \atop {\rm Hei}} (\A) = i_{\alpha \alpha \atop {\rm Hei}} V_{\A} +c_{\alpha \alpha \atop {\rm Hei}} P_{\A} + \bc_{\alpha \alpha \atop {\rm Hei} } \log l_\A + \mathcal{O}(1),
\end{equation}
where $ V_{\A}$ is here the area of $\A$, $P_{\A}$ its perimeter and $l_\A$ a characteristic length (size of the region $\A$). $i_{\alpha \alpha}$ corresponds to the QFID, while $ \bc_{\alpha \alpha}(\A)$ marks the presence of the Dirac cone;  $c_{\alpha \alpha}$ is non-universal.

$\bc_{\alpha \alpha}(\A)$ and $\bc_{\rm Hei}(\A)$ are corner contributions in the sense that they are determined solely by the sharp angles of $\A$. As illustrated in Fig.~\ref{fig2}, each corner of angle $\psi$ contributes with the corner function $a_{\alpha \alpha} (\psi)$ so that
\begin{equation}
\bc_{\alpha \alpha/{\rm Hei}}(\A) =\sum\limits_{\text{corners angles $\psi$}} a_{\alpha \alpha / {\rm Hei}} (\psi).
\end{equation}
Below, we compute the corner functions $a_{\alpha \alpha / {\rm Hei}} (\psi)$ and show their relation with the cone's winding number. We also discuss the relation with a similar scaling in the entanglement entropy.

\subsection{Generalities}
We study a system that presents only a single Dirac cone at $\vec{k}_0$ (taken to be $\vec{0}$ for simplicity of notations), with a winding number $m\neq 0$. We assume rotation invariance of the cone such that the effective low-energy Hamiltonian can be written as:
\begin{equation}
  \vec{n}(\vec{k}) = \begin{pmatrix} \text{Re}\left((k_x +\text{sgn}(m) ik_y)^{|m|}\right) \\
    \text{Im}\left( (k_x +\text{sgn}(m) ik_y)^{|m|}\right) \\  0 
  \end{pmatrix}
  \label{eq:DiracCones}
\end{equation}
close to the Dirac cone. To properly regularize the various integrals, we consider for now a square lattice, and later generalize to other forms. Note that in the $p+ip$ model, the $n_z$ component is not identically zero. Nonetheless, it vanishes faster than the other two components at the gapless point, making it irrelevant for the purpose of bipartite charge fluctuations.
%Such a behavior can be  enforced by some of the symmetries of the Hamiltonian, as is the case for example in the trivial to topological critical point in the $p+ip$ model introduced in Eq. \ref{eq:pip}. In such a case, the very symmetries that ensure topology also enforce properties on the fluctuations.

Due to the isotropy of $\vec{n}(\vec{k})$, computation of $\hat{\textbf{n}}$ at large $r$ is straightforward:
\begin{equation}
\hat{\textbf{n}}(\vec{r}) =  -\frac{(-i)^{|m|} |m|}{2 \pi} \frac{1}{r^2} ( \cos m \theta , \sin m \theta, 0) +\mathcal{O} (r^{-\frac{5}{2}}), \label{eq:CoeffDirac}
\end{equation}
where $(r, \theta)$ are the polar coordinate of $\vec{r}$. From this expression and Eq. \ref{eq:GenCorrSite}, one directly recovers the correlation functions $\C^1_\alpha$ (Eq. \ref{eq:defC1}), and  $\C^2_{\alpha \alpha}$ (Eq. \ref{eq:defC2}). \\ %\lh{the ZZ component comes from $|n_x|^2 +|n_x|^2-|n_z|^2$, which is why it is non-zero. The alpha is outside the norm: it marks the chosen Minkowski norm, and not the component.}
\begin{equation}
  \begin{split}
\C^2_{ZZ}(\vec{r}) &= \frac{q_e m^2}{16 \pi^2} r^{-4}+...\\
\C^2_{YY}(\vec{r}) &= -\C^2_{XX}(\vec{r}) = \frac{q_e m^2}{16 \pi^2} r^{-4} \cos(2 m \theta)+...
\label{eq_corr}  \end{split}
\end{equation}
The winding number of the cone can be extracted from the correlators $\C^2_{\alpha \alpha}$, either from the dominant term in $\C^2_{ZZ}$ or from the oscillation periodicity or amplitude in $\C^2_{XX}$ and $\C^2_{YY}$.

\subsection{Logarithmic contributions and corner functions}

Before discussing in length the corner function for bipartite charge fluctuations, let us briefly summarize related results for the entanglement entropy (or vNEE) in the CFT-invariant case $ m=\pm 1$. The  entanglement entropy of a subregion $\A$ takes the scaling form
\begin{equation}
\Ss(\A) = \alpha P_{\A} - \bc_{\Ss} \log l_\A + \mathcal{O}(1),
\end{equation}
where $\bc_{\Ss}$ is a corner contribution, similar but distinct from the coefficient $\bc_{\alpha \alpha}$ and $\bc_{\rm Hei}$  in Eq.~\eqref{scaling_flu}.
$\bc_{\Ss}$ is the sum over all sharp angles of $\A$,
\begin{equation}
\bc_{\Ss}=\sum\limits_{\text{corners angles $\psi$}} a_{\Ss}(\psi).
\end{equation}
where the universal corner function\citep{Bueno2015, Bueno2015-2, Elvang2015, Miao2015, Faulkner2016}  $a_\Ss$ satifies the symmetry property  $a_{\Ss}(\psi) = a_{\Ss}(2 \pi - \psi)$. Its exact analytical expression is not known, but high precision approximations can be found in Ref. \onlinecite{Helmes2016}. We compare below the corner function $a_{\Ss}(\psi)$ with the corresponding corner function $a_{ZZ}(\psi) = a_{\text{Hei}}(\psi)$ for bipartite charge fluctuations ($a_{XX/YY}(\psi)$ break rotational invariance, see Eq.~\eqref{eq_corr}).

%The logarithmic coefficient in the fluctuations is also a corner function. From Eq. \ref{eq:CoeffDirac}, we know that $||\hat{\textbf{n}}||^2_{\alpha}$ breaks the rotation invariance for $\alpha \neq Z$, and therefore depend on the orientation of $\A$. We therefore only compare the charge ($ZZ$) and total fluctuations to the CFT's entropy. 

\subsubsection{Exact corner function}
\begin{figure}
\begin{center}
\includegraphics[width=0.45\linewidth]{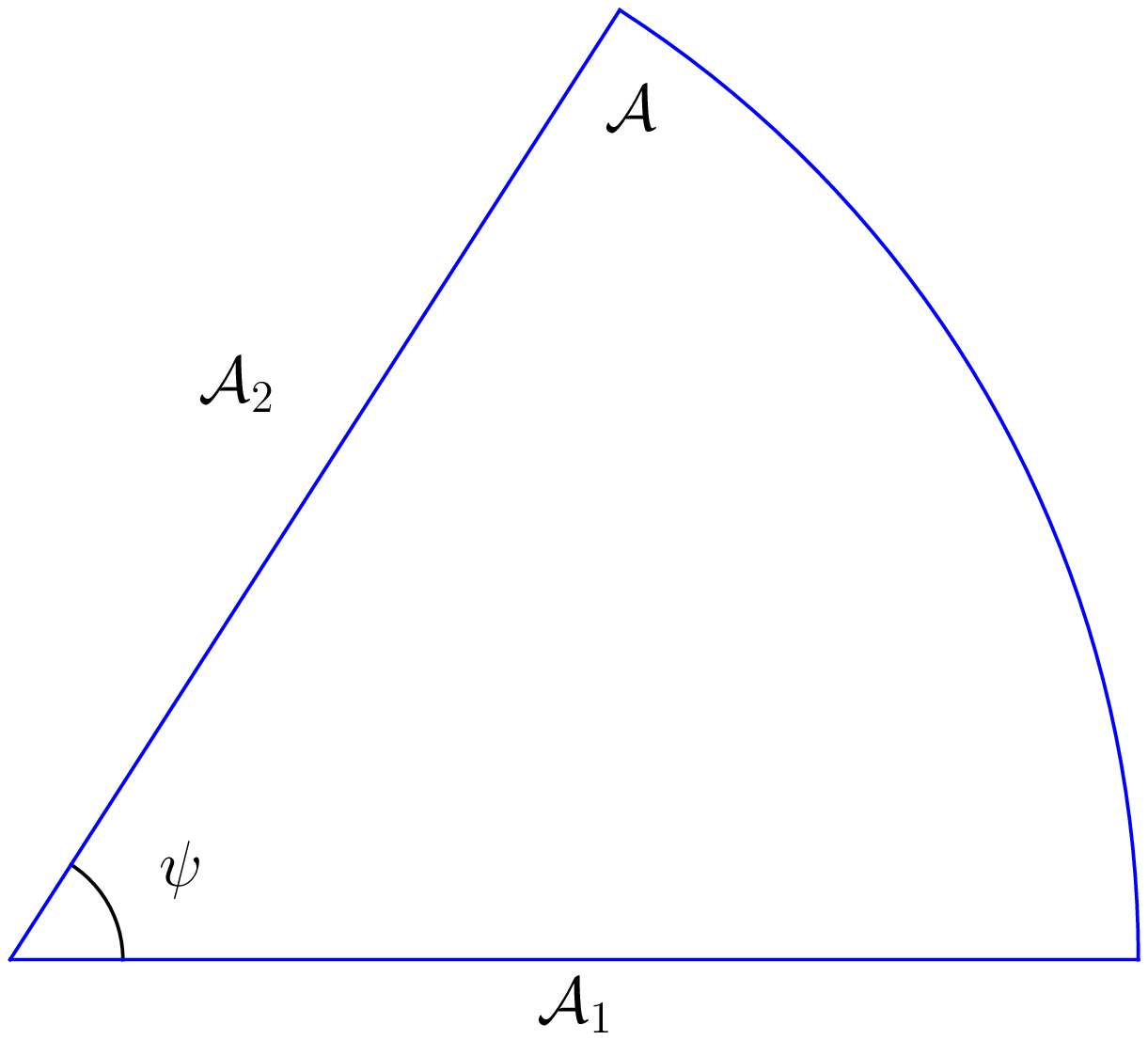}
\includegraphics[width=0.45\linewidth]{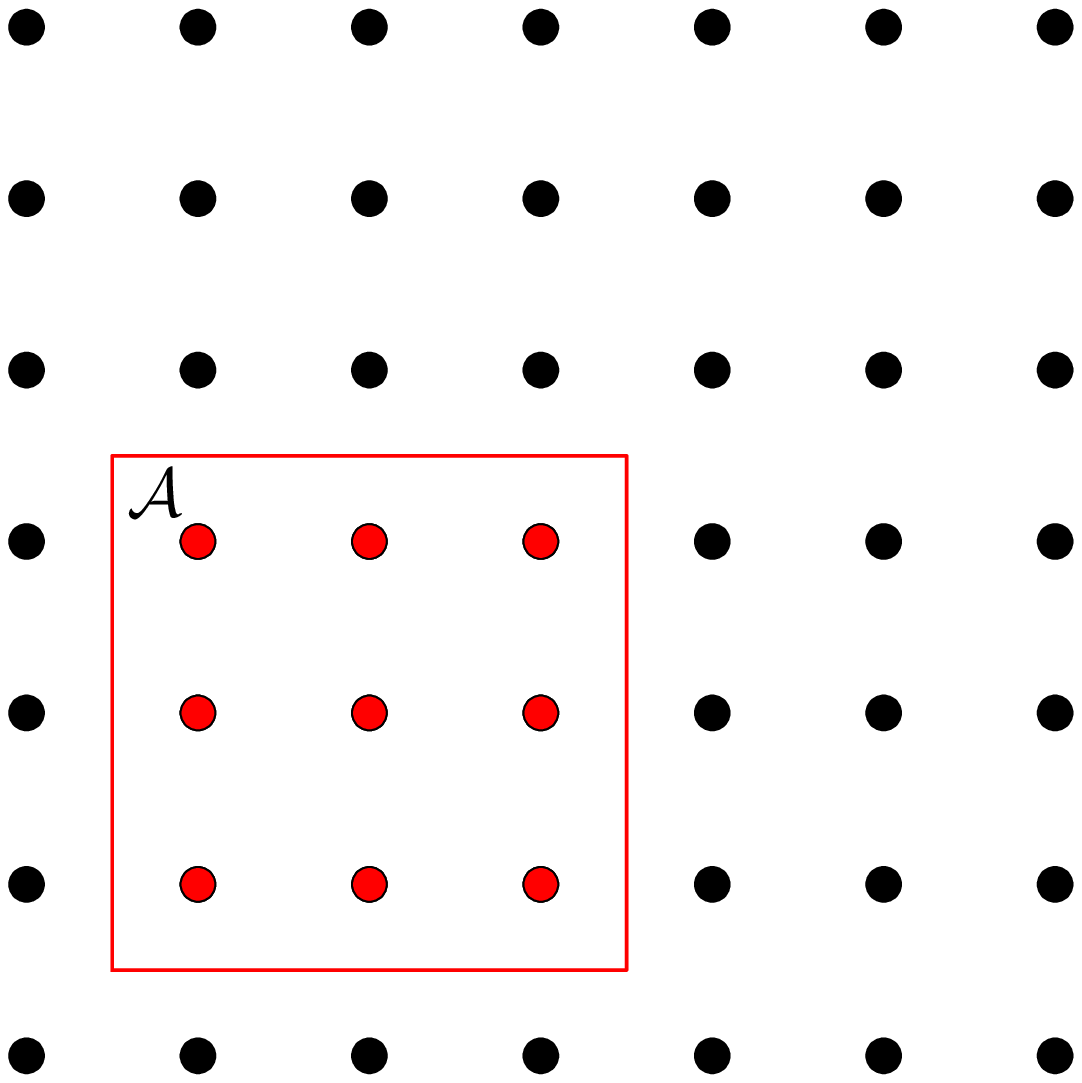}
\end{center}
\begin{center}
\includegraphics[width=0.45\linewidth]{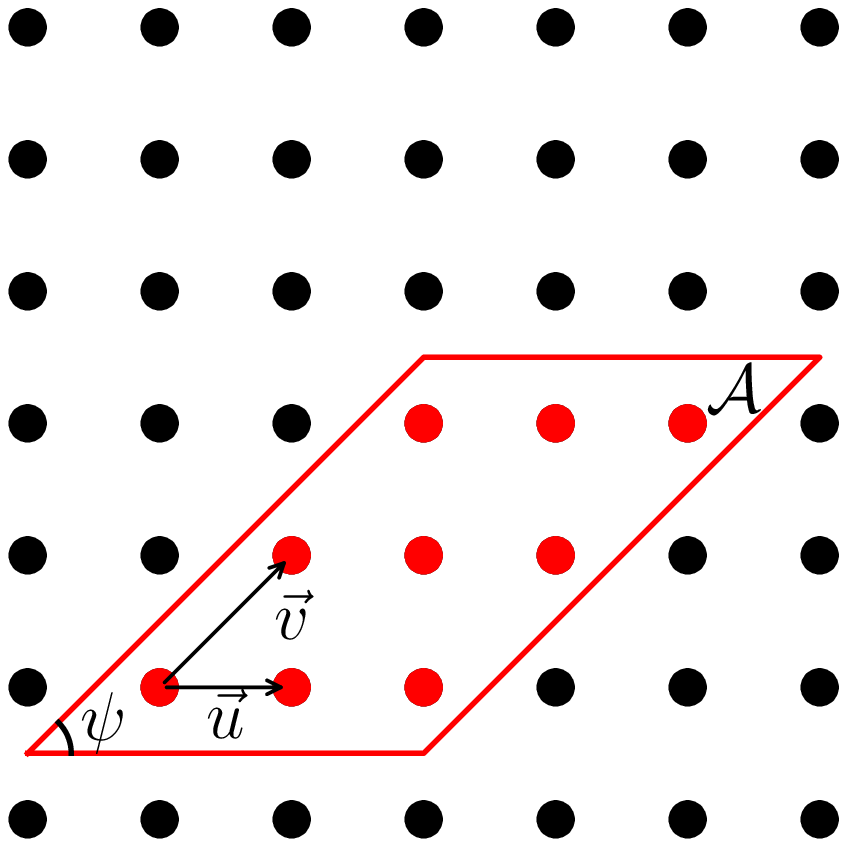}
\includegraphics[width=0.45\linewidth]{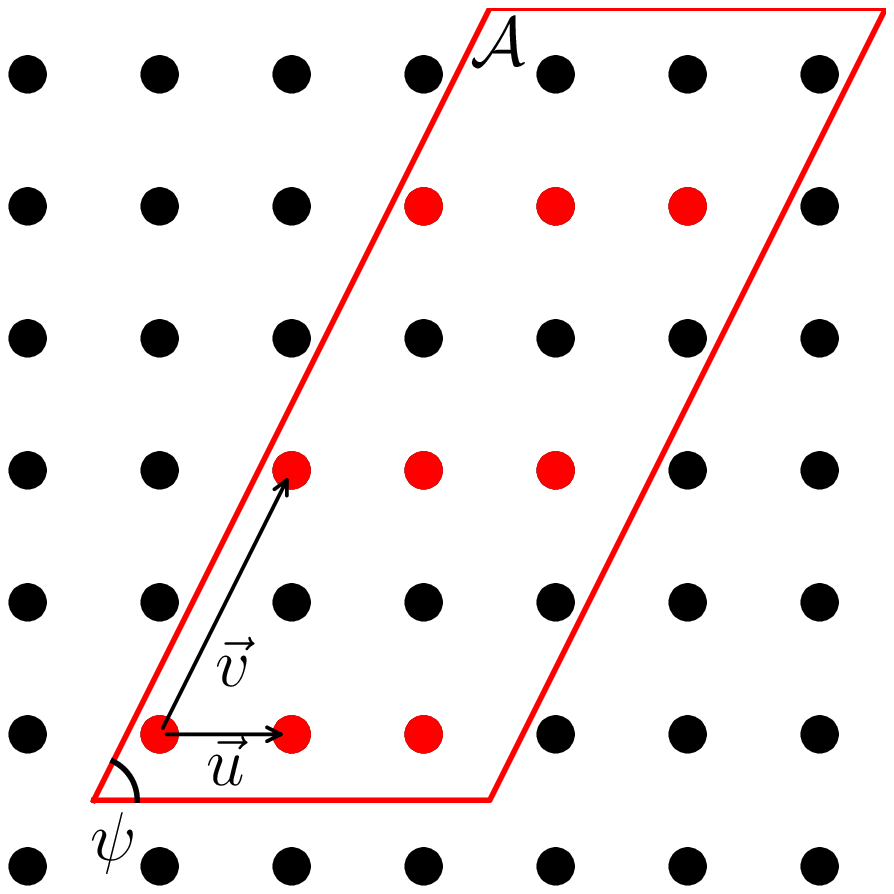}
\end{center}
\caption{Schematics of the different possibilities for the sub-system $\A$. Top left: the quadrant chosen to compute the contribution of a single corner: the obtained logarithmic coefficient correspond to  $a_{\alpha \alpha} (\psi)+2 a_{\alpha \alpha}(\frac{\pi}{2})$.  Top right: $\A$ is simply a square sub-region of the square lattice. It respects the symmetries of the model. Bottom left: we can also choose a parallelogram generated by the vectors $\vec{u}$ and $\vec{v}$. The area of the unit-cell they generate is one, and therefore the parallelogram is a complete cover of the lattice. We note $\psi$ the angle between  $\vec{u}$ and $\vec{v}$. Bottom right: another possible choice for $\vec{u}$ and $\vec{v}$. The parallelogram is still well defined, but as the area spanned by $\vec{u}$ and $\vec{v}$ is equal to $2$, all sites in $\A$ are not an integer linear combination of $\vec{u}$ and $\vec{v}$ (only those in red). A properly taken continuum limit also deals with such parallelograms.}
\label{fig2} 
\end{figure}

We focus first on computing the contribution of a single corner for the charge (ZZ) and spin-spin (Hei) fluctuations. To do so, we consider a quadrant of angle $\psi$ shown in Fig.~\ref{fig2} and compute the coefficients $\bc_{ZZ}$ and  $\bc_{\rm Hei}$ which involve three corners with angles $\psi$, $\pi/2$ and another $\pi/2$. As shown in Appendix~\ref{appen_technical}, the contributions of the different corners can be separated. We therefore obtain for a single corner of angle $\psi$
\begin{equation}
a_{ZZ}(\psi) = a_{\text{Hei}}(\psi) = \frac{q_e m^2}{32 \pi^2} \times [ 1 + (\pi-\psi) \cot \psi ], \label{exactCorner}
\end{equation}
parametrized by the winding number $m$ of the Dirac cone.

\begin{figure}
\begin{center}
\subfigure{\includegraphics[width=0.49\linewidth]{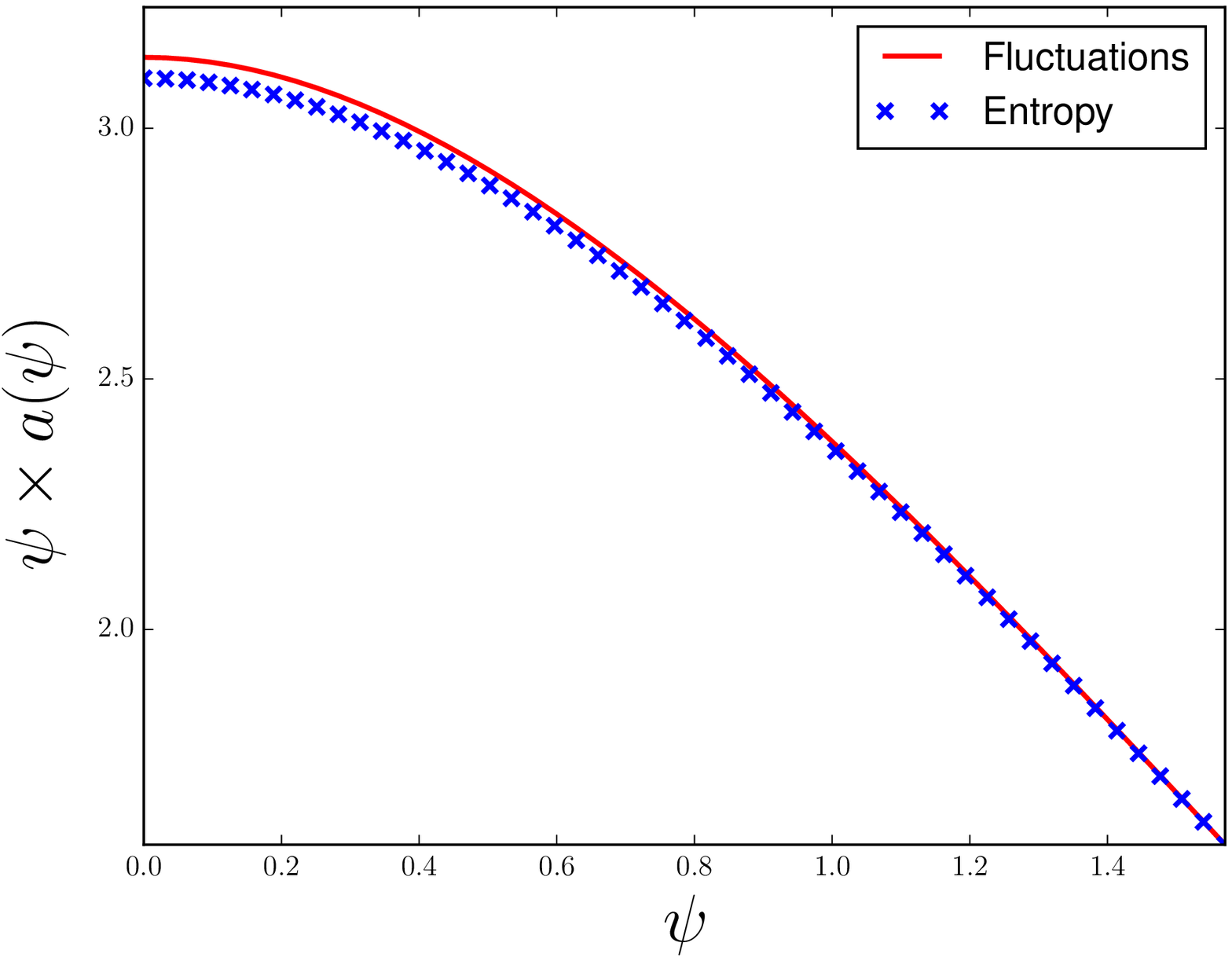}}
\subfigure{\includegraphics[width=0.49\linewidth]{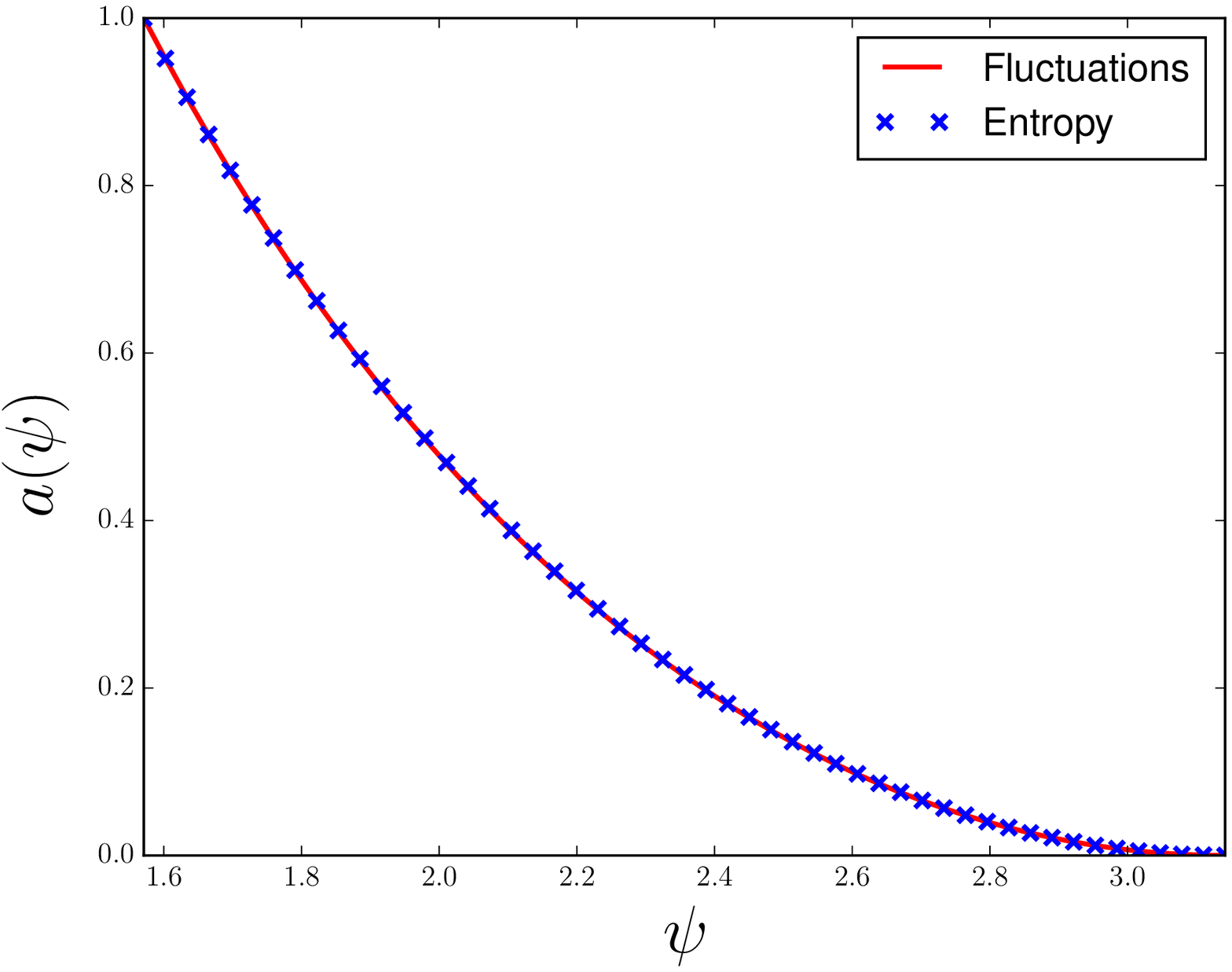}}
\end{center}
\caption{Comparison between the corner function of the BCF and of the entropy for Dirac fermions, as a function of the kink's angle. On the left graph, we represent $\psi \times a(\psi)$.}
\label{fig3}
\end{figure}

Fig.~\ref{fig3} shows the comparison between the corner function Eq~\eqref{exactCorner} in bipartite charge fluctuation and the vNEE corner function $a_{\Ss}(\psi)$ of the same model. Although close, the two functions can be clearly distinguished. As a result and in contrast with the one-dimensional case (with charge conservation), no universal ratio emerges between entanglement entropy and bipartite charge fluctuations, even when restricted to their logarithmic scaling terms, see Appendix~\ref{app:comparison} for further details.

However, the corner function Eq~\eqref{exactCorner} that we have computed coincides with vNEE corner function of another model, the Extensive Information model~\citep{Casini2009-2, Swingle2010-2, Casini2005}, which also exhibits an extensive mutual information for the vNEE, as defined in Eq.~\eqref{eq:MF}.

\subsubsection{Multiple corners: parallelograms}

We study the case where $\A$ possesses multiple corners and verify that their contributions add up separately, with the example of a parallelogram for the subregion $\A$,
\begin{equation}
 \vec{r} \in \A \text{ if } \vec{r} = r_u \vec{u} + r_v \vec{v} \text{ with } 0 \leq r_u < l_u \text{ and }0 \leq r_v < l_v.
 \end{equation} 
For $\A$ to describe a complete cover of the sublattice it encloses, the parallellogram generated by $\vec{u}$ and $\vec{v}$ must be of area $1$. For a discrete lattice, it limits the possible choices for $\vec{u}$ and $\vec{v}$, but a proper regularization allows for a direct computation at all angles. Let us introduce the notations:
$$\vec{u}= |u|( \cos \theta_u , \sin \theta_u  ), \vec{v}= |v|( \cos (\theta_u + \psi) , \sin (\theta_u + \psi) )$$
We compute the kernel $\mathcal{K}$ and obtain:
\begin{equation}
\mathcal{K} = \sum\limits_{r_u=-l_u}^{l_u} \sum\limits_{r_v=-l_v}^{l_v}  (l_u-|r_u|)(l_v-|r_v|) e^{i \vec{k}.(r_u \vec{u} + r_v \vec{v})},\label{eq:BasicPara}
\end{equation}
which, from Eq. \ref{eq:GenCorrSite}, leads to the simple expression for the bipartite fluctuations:
\begin{multline}
\F_{\alpha \alpha} (\A)=\frac{q_e V_{\A}}{4}+\\
\sum\limits_{r_u=-l_u}^{l_u} \sum\limits_{r_v=-l_v}^{l_v}  (l_u-|r_u|)(l_v-|r_v|) ||\hat{\textbf{n}}(r_u \vec{u} + r_v \vec{v})||_\alpha^2
\end{multline}
The logarithmic term can only arise from the series:
\begin{equation}
\sum\limits_{r_v=-l_v}^{l_v}  |r_u||r_v| ||\hat{\textbf{n}}(r_u \vec{u} + r_v \vec{v})||_\alpha^2,\label{eq:seriespara}
\end{equation}
whose dominant scaling term can be exactly computed (see Appendix \ref{app:Para}). We consider $l_u$ and $l_v$ to be of the same order. We introduce the corner functions $\tilde{a}_m$
\begin{align}
\tilde{a}_0(\psi)&=1+ (\frac{\pi}{2}-\psi) \cot \psi \label{eq:genCorner-1} \\
\tilde{a}_2(\psi, \theta_u)&=1 -\frac{\pi-2 \psi+\sin 2 \psi}{4\sin \psi} \cos(2 \theta_u + \psi) \label{eq:genCorner-2}\\
\tilde{a}_{2m}(\psi, \theta_u)&=\frac{1}{m^2-1} (- \cos (m \psi) \nonumber  \\
&\qquad + \frac{\sin m \psi}{m} \cot \psi) \cos (m(2 \theta_u + \psi)). \label{eq:genCorner-3}
\end{align}
We take $\tilde{a}_n=0$ for $n$ odd. The logarithmic contributions appearing for our chosen region $\A$ have the following coefficients:
\begin{align}
\bc_{ZZ}&=\bc_{\text{Hei}}=\frac{q_e m^2}{8 \pi^2} \tilde{a}_0(\psi)\\
\bc_{YY}&=-\bc_{XX}=\frac{q_e m^2}{8 \pi^2} \tilde{a}_{2m}(\psi, \theta_u)
\end{align}
The first expression coincides with the sum $\bc_{ZZ} = 2a_{ZZ}(\psi)+2a_{ZZ}(\pi-\psi)$, showing the additivity property for the different corners of the parallelogram.

\subsubsection{Lattice independence and contributions of smooth domains}
The above computations have been performed for a standard square lattice although the expression of the corner function was obtained after a continuum limit.
They can be generalized to an arbitrary lattice by again considering a single corner contribution to the logarithmic scaling term, namely $a_{ZZ}(\psi)$ and $a_{\text{Hei}}(\psi)$. The area of the Brillouin zone $A_{BZ}$ is no longer $4 \pi^2$ which renormalizes the correlators $\C^2_{\alpha \alpha}$, see Eq.~\eqref{eq_corr}, by $(4 \pi^2/A_{BZ})^2$. On the other hand, the continuum limit taken for bipartite fluctuations in Eq.~\eqref{fluctuations_form} introduces the area of the real space unit cell $A_{\vec{x}, \vec{y}}$ squared. The corner function $a_{ZZ}(\psi)$ in Eq.~\eqref{exactCorner} is therefore renormalized by the prefactor
\begin{equation}
\left(\frac{4 \pi^2}{A_{\mathcal{BZ}}}\right)^2 \times \frac{1}{A^2_{\vec{x}, \vec{y}}} = 1,
\end{equation}
and is, as a result of the lattice identity $A_{\mathcal{BZ}}A_{\vec{x}, \vec{y}} = 4 \pi^2$, unchanged. Hence, we have shown that the corner function  $a_{ZZ}(\psi)$ is valid for an arbitrary lattice and applies also for graphene or other hexagonal geometries.\\

Computing the fluctuations arising from a disk, that is to say a smooth regular subsystem without corners, is a simple check to confirm that logarithmic contributions only arise from boundary defects.
By rotation invariance, one straightforwardly obtains that:
\begin{equation}
\bc_{YY}(\A_\text{disk})=\bc_{XX}(\A_\text{disk}) = 0.
\end{equation}
The charge (and spin-spin) fluctuations do not vanish. Working in the continuum limit, we introduce the regularizing function
$$\hat{g}_\varepsilon(\vec{r}) = \frac{q_e m^2}{16 \pi^2} \frac{1}{(r^2+\varepsilon^2)^2}$$
which captures the long-range behavior of $||\hat{n}(\vec{r})||^2_{Z}$ ($\varepsilon$ is a cut-off to avoid unphysical divergence). As $\hat{g}_\varepsilon(\vec{r})||\hat{n}(\vec{r})||^{-2}_{Z}= 1 + O(r^{-1})$, the logarithmic fluctuations induced by $\hat{g}$ are the same as the one induced by $\hat{n}(\vec{r})$. From Eq. \ref{eq:GenFlucA}, one obtains for a disk $\A$:
\begin{equation}
 \int\limits_{\A^2} d\vec{r} d\vec{r}' \hat{g}_\varepsilon(\vec{r}- \vec{r'}) = \frac{q_e m^2}{4} \left(\frac{\varepsilon^2+R^2}{4 \varepsilon^2} -\frac{\sqrt{\varepsilon^4+4\varepsilon^2R^2}}{8\varepsilon^2}-\frac{1}{8}\right),
 \end{equation} 
which proves that no logarithmic term arises on the disk. It will be generally true in even space-dimensions.

\section{Two dimensions: beyond the single isotropic Dirac cone}\label{sec:Dirac-complex}
So far, we only treated minimal models that possess a single rotationally-invariant Dirac cone. In such case, the simple topological structure of the cone is easy to read in the different correlators and bipartite charge fluctuations. In this Section, we propose to go further. We start by discussing how to recover the previous results when several Dirac cones are present, as is the case in graphene or at the $\mu=0$ critical point of the $p+ip$ superconductor. We then consider the effect of anisotropies on corner functions. Though more complex, a careful study of the new corner functions  recovers the topological structure of the cones.

\subsection{Structure factor}
Due to symmetries or topological arguments, multiple Dirac cones appear in numerous condensed matter systems. The typical example is of course graphene, but  similar structures also appear at the half-filling transition point of the $p+ip$ superconductor, with Dirac cones opening at momenta $(0, \pi)$ and $(\pi, 0)$. For graphene, the two cones have opposite winding number, while the two cones have the same topological charge in the $p+ip$ superconductor.  The former is a "trivial" gapless system ( a standard mass term will typically  gap out the system into a trivial insulator) while the second has a definite topological structure (a standard mass term will typically gap it into a topological insulator). By looking at the structure factor of the fluctuations, we can identifty these two situations as they lead to different  spatial dependencies of the corner functions.\\

\subsubsection{Logarithmic contributions to the BCF}
We compute the logarithmic contributions to the BCF for several cones. We limit ourselves to the case of two Dirac cones for simplicity's sake, but results can be straightforwardly extended to any number of cones. Let $\vec{K}_\pm$ be the momenta at which the gap closes and $m_\pm$ the corresponding winding numbers. We assume that both cones are isotropic and are locally described by Eq. \ref{eq:DiracCones}. One obtains:
\begin{multline}
\hat{\textbf{n}}(\vec{r}) =  -\sum\limits_{\varepsilon=\pm} \frac{(-i)^{|m_\varepsilon|} |m_\varepsilon|}{2 \pi} \frac{1}{r^2} ( \cos (m_\varepsilon \theta+\vec{K}_\varepsilon.\vec{r}) ,\\
 \sin  (m_\varepsilon \theta+\vec{K}_\varepsilon.\vec{r}), 0) +\mathcal{O} (r^{-\frac{5}{2}}),
\end{multline}
where $\theta$ is the polar angle associated to $\vec{r}$, which in turn leads to
\begin{multline*}
\C^2_{ZZ}(\vec{r}) = \frac{q_e }{16 \pi^2} r^{-4} \left[m_+^2+ m_-^2\right. \\
\qquad+ \left.2 |m_+ m_-|\cos((m_+-m_-) \theta+ (\vec{K}_+-\vec{K}_-).\vec{r}) \right]
\end{multline*}
%\C^2_{YY}(\vec{r}) &= -\C^2_{XX}(\vec{r}) = \frac{q_e }{16 \pi^2} r^{-4}\left[m_+^2  \cos(2 m_+ \theta+2\vec{K}_+.\vec{r})\right.\\
%&\qquad+ \left.m_-^2\cos(2 m_- \theta+2\vec{K}_-.\vec{r}) + 2 |m_+ m_-|\cos((m_++m_-) \theta+(\vec{K}_++\vec{K}_-).\vec{r}) \right]
Summing the oscillating terms on a fixed $r$ contour gives a term proportional to the Bessel function $J_{m_+-m_-}(r |\vec{K}_+-\vec{K}_-|)$ and therefore leads to no contribution to the logarithmic term. The associated corner function is therefore simply given by:
\begin{equation}
\bc_{ZZ} = \frac{m^2_+ + m^2_-}{32 \pi^2} [ 1+(\pi-\psi) \cot \psi ]
\end{equation}
which generalizes to multiple cones with windings $m_j$:
\begin{equation}
\bc_{ZZ} = \frac{1}{32 \pi^2} (1+(\pi-\psi) \cot \psi) \sum\limits_j m_j^2.
\end{equation}
The logarithmic contributions of the different cones are consequently additive, similarly to entanglement entropy. Additivity of the charge fluctuations was also proven in the case of a one-dimensional Fermi surface\citep{Gioev2006}.

\subsubsection{Structure factor and relative signs}
More information can be recovered through the study of the structure factor of the BCF - or equivalently the (partial) Fourier transform of the correlation function $\mathcal{C}^2_{ZZ}$ -, defined by:
\begin{equation}
\mathcal{S}\F_{\hat{O}}(\vec{\phi}, \A) =\sum\limits_{\vec{r}, \vec{r}' \in \A} e^{i \vec{\phi}.(\vec{r}-\vec{r}')}\Braket{\hat{O}_{\vec{r}} \hat{O}_{\vec{r}'}}_c 
\end{equation}
The structure factor has similar scaling laws, with a dominating volume law of coefficient:
\begin{equation}
i_{\alpha \alpha}=\frac{q_e}{4} + \frac{q_e}{4 S} \sum\limits_{\vec{k} \in \BZ} \left(\tilde{\textbf{n}}(\vec{k}).\tilde{\textbf{n}}(\vec{k}+\vec{\phi})\right)_\alpha
\end{equation}
with $(\text{ }.\text{ })_\alpha$ the Minkowski scalar product associated to $||\text{ }||_\alpha$. Logarithmic contributions can arise if the phase is gapless. For zero-dimensional Fermi surfaces, such contributions will appear only if there exists $\vec{k}_0$ such that $\tilde{n}(\vec{k})$ is singular both at $\vec{k}_0$ and $\vec{k}_0+\vec{\phi}$. Consequently, logarithmic terms appear here only if $\vec{\phi}=\vec{0}, \pm (\vec{K}_+-\vec{K}_-)$. Indeed, one obtains:
 \begin{multline*}\text{Re}(e^{i \vec{\phi}.\vec{r}}
\C^2_{ZZ}(\vec{r})) = \frac{q_e }{16 \pi^2} r^{-4} \left[(m_+^2 + m_-^2)\cos(\vec{\phi}.\vec{r})\right. \\
\qquad+ |m_+ m_-|(\cos((m_+-m_-) \theta+ (\vec{\phi}+\vec{K}_+-\vec{K}_-).\vec{r})\\
\qquad+ \left.|m_+ m_-|\cos((m_+-m_-) \theta+ (-\vec{\phi}+\vec{K}_+-\vec{K}_-).\vec{r}) \right]
\end{multline*}
For $\vec{\phi}=\pm (\vec{K}_+-\vec{K}_-)$, the only relevant contribution is:
\begin{equation}
 \frac{q_e }{16 \pi^2} r^{-4} |m_+ m_-|\cos((m_+-m_-) \theta).
\end{equation}
As discussed in Appendix \ref{app:Para}, if $m_+-m_-$ is odd, no logarithmic term appears in bipartite charge fluctuations. On the other hand, if it is even, a logarithmic term is present with prefactor
\begin{equation}
 \frac{q_e |m_+ m_-| }{16 \pi^2} \tilde{a}_{m_+-m_-}(\phi, \theta_u),
\end{equation}
with $\tilde{a}$ defined in Eq. \ref{eq:genCorner-1}-\ref{eq:genCorner-3}. While in the thermodynamic limit the logarithmic terms only appear precisely at $\vec{\phi}=\vec{0}, \pm (\vec{K}_+-\vec{K}_-)$, finite-size effects induce logarithmic corrections when $ ||-\vec{\phi}+\vec{K}_+-\vec{K}_-||^2 \ll  A_\A^{-1}$, where $A_\A$ is the area of the larger subregion $\A$ considered.

\subsection{Anisotropies}
Introduction of anisotropies in the energy dispersion is necessary if one wants to analyze the response of real materials. On the theory side, it is also interesting as rotational invariance is broken as well as conformal symmetry. The simple analytical structure of bipartite charge fluctuations allows for the computation of the induced anisotropic corner functions. To simplify notations, we only consider models based on a square lattice, and cones with winding number $\pm 1$.

We consider a general case where \begin{equation}
\vec{n}_x(\vec{k}) \approx \Delta^x_x k_x + \Delta^y_x k_y, \text{ } \vec{n}_y(\vec{k}) \approx \Delta^x_y k_x + \Delta^y_y k_y
\end{equation} close to the Dirac cone. $n_z(\vec{k})$ is still assumed to be of higher order. The $p+ip$ superconductor considered in Eq. \ref{eq:pip} reduces to such a low-energy theory for $|\mu|=4t$ and $|\Delta_x|\neq |\Delta_y|$. Let us define the transformation:
\begin{equation*}
R = \matd{\Delta_x^x}{\Delta_x^y}{\Delta_y^x}{\Delta_y^y}, \qquad R^{-1} =\frac{1}{J} \matd{\Delta_y^y}{-\Delta_x^y}{-\Delta_y^x}{\Delta_x^x},
\end{equation*}
with $J=\text{det}(R)=\Delta^x_x \Delta^y_y -\Delta^x_y \Delta^y_x$. The winding number of the Dirac cone is  $ \text{sign} (J).$
When it cancels, the winding is indeed $0$ and there are no logarithmic contributions. The logarithmic contribution to the BCF is captured by the test function:
\begin{equation}
h_R(\vec{k}) = \frac{\Delta^x_x k_x + \Delta^y_x k_y + i (\Delta^x_y k_x + \Delta^y_y k_y)}{|\Delta^x_x k_x + \Delta^y_x k_y + i (\Delta^x_y k_x + \Delta^y_y k_y)|} h^{\text{reg}}_R(\vec{k}),
\end{equation}
where $h^{\text{reg}}$ is a smooth, arbitrary cut-off function with $h^{\text{reg}}(0)=1$. We then compute $\hat{h}_R$, the Fourier transform of $h_R$. Taking $h^{\text{reg}}_R$ such that $h^{\text{reg}}_R(\vec{k}'))=1-|\vec{k}'|$, one obtains
\begin{equation}
\hat{h}_R(\vec{r}) \approx\frac{e^{i(\theta_{^tR^{-1}\vec{r}}-\frac{\pi}{2})}}{2 \pi |J|} \frac{1}{|^tR^{-1} \vec{r}|^2} + ...
\end{equation}
$\theta_{^tR^{-1}\vec{r}}$ is the polar coordinate of the vector $^tR^{-1}\vec{r}$. If $R$ is an orthogonal transformation, the logarithmic coefficient is consequently not affected by the transformation. Indeed, such a transformation is equivalent to a simple change of basis.

More general transformations deform the corner functions as they locally change the metric and angles. Moreover, as anisotropies appear, the corner function becomes also function of the direction of the region $\A$. %The most convenient form to recover the coefficient is:
%\begin{multline*}
%\sum\limits_{\vec{r}_1, \vec{r}_2 \in \A}\text{ } \iint\limits_{\vec{k}\in \mathcal{BZ}} \frac{d \vec{k}}{A_{BZ}} e^{i \vec{k}.(\vec{r}_1-\vec{r}_2)} h_R(\vec{k}) \\
%=\sum\limits_{\vec{r}_1, \vec{r}_2 \in ^tR^{-1}(\A)} \text{ }\iint\limits_{\vec{k}\in R(\mathcal{BZ})} \frac{d \vec{k}}{|J|A_{BZ}} e^{i \vec{k}.(\vec{r}_1-\vec{r}_2)} h_R(R^{-1}\vec{k})
%\end{multline*}
%\begin{figure}[h]
%\begin{center}
%\includegraphics[width=0.8\linewidth]{CoefficientParaAnis.eps}
%\end{center}
%\caption{Extracted logarithmic contribution in the $p+ip$ superconductor for several values of $\Delta^y$ and $\Delta^x=t$ at the critical point $\mu=-4t$. The long range behavior is independent of the anisotropy when considering an aligned square sub-system $\A$. $l_\A$ is here the length of the side of $\A$. We recover the predicted coefficient. \lh{Typo dans les ordonnees du graphes}}
%\label{fig:III-CoefficientParaAnis}
%\end{figure}

Deformations of the cone are equivalent to deformations of the region $\A$. Note that the transformation $R$ cannot make new angles appear: logarithmic contributions still arise from the original angles, whose amplitudes are renormalized. We give as example an analytical formula when $R$ is a simple anisotropic dilatation. It corresponds to a cone where the two velocities in the $x$ and $y$ directions differ.\\

We have then: $\Delta^x_y=\Delta^y_x=0$ and $\Delta^y_y = \alpha \Delta^x_x$, $\alpha>0$. Let us consider $\A$ a parallelogram define by $\vec{u} = |u|(\cos \theta_u, \sin \theta_u)$ and $\vec{v} = |v|(\cos (\theta_u + \psi), \sin( \theta_u + \psi))$, with $\psi$ in $[0, \frac{\pi}{2}]$, represented in Fig. \ref{fig2}. Then the angle between $^tR^{-1}\vec{u} $ and $^tR^{-1}\vec{v}$ is still in the first quadrant and given by:
\begin{widetext}
\begin{equation}
\tilde{\psi}(\psi, \theta_u) = \arcsin \frac{\alpha \sin \psi}{\sqrt{(1+(\alpha^2-1) \sin^2 \theta_u)(1+(\alpha^2-1) \sin^2(\theta_u+\psi))}}
\end{equation}
\end{widetext}
and the associated corner function is simply $a_\alpha(\psi, \theta_u) = a(\tilde{\psi})$ with $a$ the corner function for the isotropic cone. Its coefficient is invariant. $\psi$ in $[\frac{\pi}{2}, \pi]$ is obtained by symmetry. In particular, for $\vec{u}=\vec{e}_x$ and $\vec{v}=\vec{e}_y$, the logarithmic coefficient is not affected.\\%, as shown in Figure \ref{fig:III-CoefficientParaAnis}.\\

%A pure deformation: $R = \matd{1}{\sin \alpha}{0}{\cos \alpha}$. Similar computations lead to a renormalization of $\psi$ in $[0, \frac{\pi}{2}]$
%\begin{widetext}
%\begin{equation}
%\tilde{\psi}(\psi, \theta) = \arcsin \frac{\sin \psi \cos \alpha}{\sqrt{(1-\sin \alpha \sin 2 \theta)(1-\sin \alpha \sin (2 \theta + 2\psi))}}
%\end{equation}
%\end{widetext}
%and similarly the associated corner function is $a_\alpha(\psi, \theta) = a(\tilde{\psi})$, while the prefactor is not affected.

\section{Higher dimensions: Weyl semi-metals}\label{sec:HighD}
Our previous results can be extended to higher dimensions, and in particular to Weyl semi-metals. Bipartite charge fluctuations can be used to directly identify the chiral charges of Weyl nodes. In this Section, we investigate bipartite charge fluctuations for isotropic Weyl nodes and give formulas for three-dimensional generalizations of corner functions. Weyl points translate into non-resolvable points in the flattened Hamiltonian $\tilde{\textbf{n}}(\vec{k})$, which are responsible for logarithmic contributions to bipartite fluctuations. The general scaling law for bipartite charge fluctuations has the form
\begin{equation}\label{fluctuations3d}
\F_{\alpha  \alpha}(\A) =  i_{\alpha  \alpha} V_\A + c_1 A_\A + c_2 l_\A  + \bc_{\alpha \alpha} \ln l_\A + \mathcal{O}(1),
\end{equation}
with $V_\A$ the volume, $A_\A$ the surface and $l_\A$ a characteristic length of $\A$. $i_{\alpha, \alpha}$ is still the quantum Fisher information density. Note that in three dimensions, logarithmic contributions can also arise from smooth (curved) entangling surfaces.

We focus below on the most simple case of a single isotropic Weyl point with chirality $m = \pm 1$ contributing to bipartite spin-spin (Hei) fluctuations. We derive in particular closed-form expressions for a set of corner functions. We also briefly discuss the generalization to Weyl points with a higher charge.

%Finally, we introduce a way to generalize all the results in this paper to models in higher-dimensions or with a larger number of bands: as long as the effective low-energy Hamiltonian can be recast in term of matrices representing the $SO(M)$ algebra, similar properties will hold.\\

% \subsection{Isotropic Weyl points of charge $\pm 1$}
%We study in this Section the contribution of a single isotropic Weyl point to bipartite spin-spin (Hei) fluctuations. We derive closed-form expressions for corner functions in the case of chirality  $\pm 1$. Larger chirality are discussed in the next Section.

%In this Section, we propose to study the contribution of a single Weyl point to the Heisenberg fluctuations, starting with isotropic points of chirality  $\pm 1$, and give some close forms for corner functions. Then we discuss the case of larger chirality, and as in the previous Section, study the typical signature of the presence of several Weyl points, as these usually come in (at least) pairs. We assume an underlying cubic lattice for simplicity.

\subsection{Model and universal correlation functions}
Let us assume an isometric ($v_x=v_y=v_z=v_F$ in Eq. \ref{eq:Weyl2}) Weyl point centered at the momentum $\vec{k}=\vec{0}$, with the  low-energy Hamiltonian
\begin{equation}
\vec{n}(\vec{k}) = v_F D \vec{k}, \text{ with } D=\text{diag}(\pm 1, \pm 1, \pm 1), \label{eq:HamWP}
\end{equation}
with $v_F$ the Fermi velocity. For the specific Weyl model of Eq.~\eqref{Weyl_model}, one has $v_F=|2t_x|=|2t_y|=\sqrt{4t^2-(B_z+4t)^2}$. The chiral charge of the Weyl point is  $\text{det}(D)$. Without loss of generalities, we can fix two of the components of $D$, and take  
\begin{equation}
\vec{n}(\vec{k}) = v_F D_{\pm 1}\vec{k}, \text{ with } D_{\pm 1}=\text{diag}(1, \pm 1, 1).
\end{equation}
Either taking advantage of the rotational invariance or by using the plane wave decomposition in spherical harmonics and spherical Bessel functions (see Appendix \ref{app:WPCorr} for a detailed computation), one obtains the dominant contribution to $\hat{\textbf{n}}$:
\begin{equation}\label{eq_nr}
\hat{\textbf{n}}(\vec{r}) = \frac{i}{\pi^2} D_{\pm 1} \frac{\vec{r}}{r^4} + \mathcal{O}(r^{-4})
\end{equation}
As in two dimensions, this contribution is universal: microscopic details of the theory will only affect less relevant terms as long as the WP can be described by the Hamiltonian in Eq. \ref{eq:HamWP}.

\subsection{Logarithmic term in fluctuations}
We focus on the Heisenberg spin-spin fluctuations $\F_{\text{Hei}}$ since the other charge fluctuations depend on the orientation of $\A$. While logarithmic contributions are present at the same order, contrarily to the two-dimensional case, they will be present even for a spherical subregion $\A$. We derive the contributions to the logarithmic term $\bc_{\text{Hei}}$ in Eq.~\eqref{fluctuations3d} for the single Weyl node of Eq.~\eqref{eq:HamWP} and for different geometries of  $\A$.
 We use the regularizing function $$ \frac{q_e }{4\pi^4} \frac{1}{(r^2+\varepsilon^2)^3}$$
which has the same behaviour at large $r$ as $(q_e/4)  ||\hat{\textbf{n}}(\vec{r})||^2$, 
see Appendix \ref{app:WPLog} for detailed computations.

When the sub-region $\A$ is a sphere, we obtain
\begin{equation}
\bc_{\text{Hei}} = \frac{q_e }{12 \pi^2}, \bc_{XX} = \bc_{YY} = \bc_{ZZ}=\frac{q_e }{36 \pi^2}
\end{equation}
For $\A$ a cylinder of radius $R$ and length $l$, we obtain:
\begin{equation}
\bc_{\text{Hei}}=  \frac{q_e}{64 \pi^2} \frac{l}{R}
\end{equation}
Then, two types of singularities in the region $\A$ may lead to anomalous contributions. The presence of (infinite) wedges lead to a logarithmic term:\begin{equation}
\bc_{\text{Hei}}=  -\frac{1}{48 \pi^4}(1+(\pi-\psi) \cot \psi),
\end{equation}
where $\psi$ is the angle formed by the wedge. Finally, the presence of singular points forming a cone leads to universal unusual terms of the form:
\begin{equation}
 -\frac{q_e}{128 \pi^2} \cos \psi \cot \psi \log^2 l
\end{equation}

These results are coherent with the form obtained for the entropy in $3+1D$ CFTs\citep{Solodukhin2008,Solodukhin2010, Casini2010, Huerta2012, Myers2012, Miao2014} and we conjecture that fluctuations follow the same geometry-imposed rules.

\subsection{Multiple contributions: parallelepipeds}
As a general rule, Eq. \ref{eq:App-Entangling} indicates that the contributions of the different corners on the boundary are additive if they lie far enough from each other. On the other hand, it is possible for them to interfere in more complex geometries. The typical example we present here is the case of a parallelepiped. Indeed, the different wedges that appear in such a geometry are not independent: the three intersecting wedges forming 3D corners interfere and lead to a specific response which cannot be reduced to the sum of its parts.

The method used for computing the integral is a generalization to three dimensions of the method presented in App \ref{app:Para}. For a parallepiped generated by the three vectors $(\vec{e}_x, \tan \psi~\vec{e}_x+ \vec{e}_y, \vec{e}_z)$, one obtains a contribution of the form
\begin{equation}
-\frac{1}{4 \pi^4} \left[ 1+(\frac{\pi}{2}-\psi) \cot \psi \right]
\end{equation}
The different wedges contributions are no longer additive as wedges are not infinite and cannot be considered independently.\\

We conclude this Section by mentioning the case of Weyl nodes with a higher chirality such as the low-energy form
\begin{equation}
  \tilde{n}_{m}(\vec{k})= \begin{pmatrix}
    \sin \theta_k \cos (m \psi_k) \\ \sin \theta_k \sin (m \psi_k) \\ \cos \theta_k 
  \end{pmatrix},
\end{equation}
for the flattened Hamiltonian, corresponding to a chiral charge $m$. The same calculation as for $m=\pm 1$ can be reproduced by using the spherical harmonics decomposition of the Fourier transform discussed in Appendix~\ref{app:WPCorr}. The resulting expressions for correlation functions and bipartite functions are nevertheless cumbersome and shall not be given here.

\section{Conclusion}\label{sec:Conclusion}
Bipartite charge fluctuations and the long-range behaviour of two- and four-body correlators are related to the non-analyticies of the Hamiltonian at the Fermi surface. Semi-metals, which have a set of Fermi points instead of a surface, exhibit topological markers such as non-trivial winding numbers around these non-analytical Fermi points. We showed that charge and spin correlators present asymptotic behaviour at large distance controlled by the vicinity of the Fermi points with coefficients directly related to winding numbers. They thus provide an alternative probe for topological invariants characterizing isolated gapless nodes. Moreover, these correlators translate into bipartite charge fluctuations with subleading logarithmic scaling terms which also depend on the winding numbers and can be measured. We obtained that the logarithmic term results from additive corner contributions for which we derived a series of analytical expressions. Although the von Neumann entanglement entropy also exhibits a logarithmic scaling term with a corner structure, its corner functions are found to be quantitatively different from those of bipartite charge fluctuations.

In cases where there are many gapless Fermi points, such as Dirac cones or Weyl nodes, the structure factor associated to bipartite charge fluctuations recovers the topological charges (winding numbers) of the different nodes and therefore distinguishes a non-topological graphene-like structure, with two opposite topological charges, from a topogical $p+ip$ superconductor-like structure (at half-filling), with two identical topological charges. Within these two models, the structure factor thus probes the topological character of the system. Although our study focused on two-dimensional topological superconductors, it can be generalized to topological insulators, such as Haldane's honeycomb model~\citep{Haldane1988}. Our results also extend to situations where rotational symmetry is broken.

Finally, let us mention that our results can also be extended to nodeless models in higher dimensions and/or with a larger number of relevant bands. The low-energy Hamiltonian in $d$ dimensions must take the particular form
\begin{equation*}
H = \sum\limits_{\vec{k} \in \BZ} \Psi^\dagger_{\vec{k}} \, \vec{n}(\vec{k}).\vec{\gamma} \, \Psi_{\vec{k}},
\end{equation*}
where the  matrices $\vec{\gamma}$ satisfy  the Clifford algebra
\begin{equation}
\{ \gamma_j, \gamma_k \} = 2\delta_{j,k},
\end{equation}
and form an irreducible representation of $SO(d)$. The number of components of the vector $\vec{n}(\vec{k})$ matches the number of independent matrices $\gamma_j$. A simple example for $SO(5)$ is provided by the set of matrices
\begin{align}
\gamma_0&=\sigma^z\otimes I_2,~~ \gamma_1=\sigma^y\otimes\sigma^x,~~  \gamma_2=\sigma^y\otimes\sigma^y,\nonumber \\
 \gamma_3&=\sigma^y\otimes\sigma^z, ~~ \gamma_5=\sigma^x\otimes I_2
\end{align}
describing a subspace of four bands. For the particular isotropic case where $\vec{n}(\vec{k}) \propto \vec{k}$, corresponding to a sort of higher-dimensional Weyl node, the correlation function can be computed
\begin{align}
\hat{n}(\vec{r})&=i\frac{\Gamma(\frac{1+d}{2})}{\sqrt{\pi}^{d+1}} \frac{\vec{r}}{r^{d+1}} +  \mathcal{O}(\frac{1}{r^{d+1}})
\end{align}
where $\Gamma$ is the Gamma function, and used to determined bipartite charge fluctuations of the model. Also here, the emergence of these universal long-range properties are signatures of the presence of gapless points with a specific low-energy form.\\

%Finally, we discuss the extension of these results to higher-dimensional or multiband systems. Most of the results derived in this paper can be straightforwardly extended to Hamiltonians in more than three dimensions, or with more than two-bands at low-energy, if they can be expressed as
%\begin{equation*}
%\frac{q_e}{2} \sum\limits_{\vec{k} \in \BZ} \Psi^\dagger_{\vec{k}} \vec{n}(\vec{k}).\vec{\gamma} \Psi_{\vec{k}},
%\end{equation*}
%where $\vec{\gamma}$ form a unitary representation of the Clifford algebra.

\paragraph*{Acknowledgements.} This work has benefited from useful discussions with N. Regnault, W. Witczak-Krempa, J. Bardarson and A. Mesaros. We acknowledge financial support from the PALM Labex, Paris-Saclay, Grant No. ANR-10-LABX-0039, the ERC Starting Grant No. 679722 and from the German Science Foundation (DFG) FOR2414 and from ANR, BOCA. We acknowledge discussions at the Center of Recherhes de Mathematiques de U. Montreal, related to the workshop on entanglement, integrability, topology in many-body quantum systems. We also acknowledge discussions at CIFAR meetings in Canada.

\newpage 
\appendix
\section{Bogoliubov formalism and observables} \label{app:bogo}
In this Appendix, we  give some more details on the Bogoliubov formalism.\\
The Hamiltonian (\ref{eq:HamGen}) can be diagonalized by the following Bogoliubov transform. We define
$(E(\vec{k}), \theta_{\vec{k}}, \phi_{\vec{k}})$ the spherical coordinates of the vector $\vec{n}_k$, the diagonalizing matrix $P_{\vec{k}}$ and the Bogoliubov-de Gennes spinor $\Upsilon_{\vec{k}}$:
\begin{equation}
P_{\vec{k}}=\matd{\cos(\frac{\theta_{\vec{k}}}{2})}{e^{-i\phi_{\vec{k}}}\sin(\frac{\theta_{\vec{k}}}{2})}{-e^{i\phi_{\vec{k}}}\sin(\frac{\theta_{\vec{k}}}{2})}{\cos(\frac{\theta_{\vec{k}}}{2})}\qquad \Upsilon_{\vec{k}}= P_{\vec{k}} \Psi_{\vec{k}},
\end{equation}
where $\Upsilon_{\vec{k}}^\dagger= \vecth{\eta^\dagger_{\vec{k},+}}{\eta_{{\vec{k}},-}}$ (for superconducting spinors, $\eta_{{\vec{k}},-}=\eta^\dagger_{-\vec{k}}$). The Hamiltonian is now
\begin{equation}
H=\frac{q_e}{2} \sum\limits_{\vec{k} \in \BZ} =E(k)(\eta^\dagger_{\vec{k}, +} \eta_{\vec{k}, +}+  \eta^\dagger_{\vec{k}, -} \eta_{\vec{k}, -})
\end{equation}
For gapped systems ($E(k)>c \in \mathbb{R}^{+*}$) the ground state $\Ket{0_{\eta}}$ cancels all $\eta_{\vec{k}, \pm}$ operators. For gapless systems, some quasi-particles may have strictly zero energy, leading to a degeneracy in the ground state. This degeneracy will not affect our results in the thermodynamic limit, and we always compute the average in $\Ket{0_{\eta}}$. \\

To compute the fluctuations in Eq. \ref{eq:GenCorrSite}, we define $\hat{O}_{\vec{r}, \alpha, \beta}=c^\dagger_{\vec{r}, \alpha} c_{\vec{r}, \beta}$, with $\alpha$ and $\beta$ indexing the species.
\begin{multline}
\Braket{\hat{O}_{\vec{r}, \alpha, \beta} \hat{O}_{\vec{r}', \alpha', \beta'}}_c = \frac{1}{S^2} \sum\limits_{\vec{k}, \vec{q} \in \BZ}  \mathcal{K}_{\C^2}(\vec{k}-\vec{q}, \vec{r}-\vec{r'})\\  \mathcal{G}_{\beta, \alpha'}(\vec{q})(\delta_{\alpha, \beta'}-\mathcal{G}_{\beta', \alpha}(\vec{k}))\label{eq:GreenRep}
\end{multline}
with $\mathcal{K}_{\C^2}(\vec{k}, \vec{r})= e^{i \vec{k}.\vec{r}}$ and $\mathcal{G}_{\alpha, \beta}(\vec{k})=\Braket{c_{\vec{k}, \alpha} c^\dagger_{\vec{k}, \beta}}$ the Green's functions. $S$ is the total number of unit-cells in the system.\\
The expressions of the fluctuations can then be obtained after some algebra, and are summarized in Table \ref{tab:InsDBF}

\begin{table}
\begin{center}
\begin{tabular}{|c|c|}
\hline Term $\alpha \beta$ & Integral form $g_{\alpha \beta}$\\
\hline ZZ & $1-\cos \theta_{\vec{k}} \cos \theta_{\vec{q}} + \cos(\phi_{\vec{k}}-\phi_{\vec{q}}) \sin \theta_{\vec{k}} \sin \theta_q $ \\
\hline XX & $1+\cos \theta_{\vec{k}} \cos \theta_{\vec{q}} - \cos(\phi_{\vec{k}} + \phi_{\vec{q}}) \sin \theta_{\vec{k}} \sin \theta_{\vec{q}} $ \\
\hline YY & $1+\cos \theta_{\vec{k}} \cos \theta_{\vec{q}} + \cos(\phi_{\vec{k}} + \phi_{\vec{q}}) \sin \theta_{\vec{k}} \sin \theta_{\vec{q}} $\\
\hline XZ & $-4 \cos \phi_{\vec{q}} \sin \theta_{\vec{q}} \cos \theta_{\vec{k}}$ \\
\hline YZ & $-4 \sin \phi_{\vec{q}} \sin \theta_{\vec{q}} \cos \theta_{\vec{k}}$ \\
\hline XY & $- 2\sin (\phi_{\vec{k}} + \phi_{\vec{q}}) \sin \theta_{\vec{k}} \sin \theta_{\vec{q}}$ \\ \hline
\end{tabular}
\end{center}
\caption{Expressions for the two-point correlator $\C^2$ and the bipartite fluctuations for an arbitrary polarization for the topological insulators. $\theta_{\vec{k}}$ and $\phi_{\vec{k}}$ are the spherical coordinates of $\tilde{\textbf{n}} (\vec{k})$ We take the following convention: if $\hat{O}_{\vec{m}}= \frac{q_e}{2}\Psi^\dagger_j \vec{m}.\vec{\sigma} \Psi_j$, avec $\vec{m}=(m_x, m_y, m_z) \in \mathbb{R}^3$, then the associated correlations or fluctuations are given by: $\mathcal{F}_{\hat{O}_{\vec{m}}}(\A) = \frac{q_e }{4S^2} \sum\limits_{\alpha\leq \beta=x,y,z} m_\alpha m_\beta \sum\limits_{{\vec{k}},{\vec{q}} \in \mathcal{BZ}} \mathcal{K}(\vec{k}-\vec{q}, \vec{r}) g_{\alpha \beta}({\vec{k}},{\vec{q}})$. The Kernel $\mathcal{K}(\vec{k}, \A)$ is simply $e^{i \vec{k}.\vec{r}}$.}
\label{tab:InsDBF}
\end{table}

\section{Scaling laws and Sobolev spaces}\label{app:Fej}
\subsection{Singularities and Sobolev spaces}
The scaling of the Fourier transform of a function $g$ is directly related to its non-analyticities. Let us start with a one-dimensional example. $\hat{g}$ is the Fourier transform of a periodic function $g$. If $g$ is $p$-differentiable such that:
\begin{equation}
\left.\begin{array}{c}
\text{$\forall 0\leq j<p,\text{ }g^{(j)}$ is continuous}\\
\text{$g^{(p)}$ is continuous by part }
\end{array}\right\rbrace \text{then $\tilde{g}(r) = O(r^{-(p+1)})$} \label{eq:1DScalingFT}
\end{equation}
We then recover instantly some well-known results by applying these results to $\tilde{n}(\vec{k})$ and the correlators $\mathcal{C}^{1/2}$ (actually all correlators due to Wick's theorem).
\begin{itemize}
\item For a gapped system, the energy $E(k)$ never cancels. In the absence of long-range term in the Hamiltonian, $\tilde{\textbf{n}}(k)$ is infinitely differentiable. Its Fourier transform therefore decreases faster than any power-law, corresponding to the exponential decay of correlations.
\item For gapless systems, the gap cancels.  $\tilde{\textbf{n}}(k)$ may be discontinuous at the gap closing points, which leads to a decay of $\hat{\textbf{n}}(r)$ of order $O(r^{-1})$. 
\end{itemize} 
For multi-dimensional Fourier transform, the scaling of the FT will depend both on the dimension of the singular manifold (here the Fermi surface), and on the order of the singularities. The proper mathematical notion for classifying the correlation functions is the concept of Sobolev spaces\citep{Giovanni}. We here briefly introduce this notion, and some expected results, and refer to standard textbooks for a complete description.\\
We define the Hilbert-Sobolev space $\mathbb{H}^m(\mathbb{T}^d)$ on the $d$-dimensional torus $\mathbb{T}^d$ as the space of the functions on $\mathbb{T}^d$ such that:
\begin{equation}
g \in H^m(\mathbb{T}^d) \Leftrightarrow \sum\limits_{\vec{r} \in \mathbb{Z}^d} |\hat{g}(\vec{r})|^2(1+|\vec{r}|^2)^{m} <+\infty
\end{equation}
Physically, the functions we are interested in will only present non-analyticities on the Fermi surface, \textit{i.e.} the manifold consisting in the vectors $ \vec{k} \in \BZ$ verifying $E(\vec{k})=0$, of dimension $d_F$\footnote{To be more precise, the Fermi Surface can consist in a set of such connected manifolds. The contributions of the different manifolds are independent.}.  For example, free non-interacting fermions in $d$ dimensions generally have a $d-1$ dimensional Fermi surface, while  Dirac  and Weyl semi-metals have a zero-dimensional (point-like) manifold.\\
As a generic rule, a function $g$ defined on $\mathbb{T}^d$ discontinuous on a $d_F$-dimensional manifold is in $H^{\frac{d-d_F}{2}-\varepsilon}(\mathbb{T}^d)$ but not necessarily in $H^{\frac{d-d_F}{2}}(\mathbb{T}^d)$.

\subsection{Kernel properties for the bipartite fluctuations}
It is convenient to  work in a simple geometry to get a better idea of the different scaling terms that may appear in the bipartite fluctuations, or in the correlations due to the kernel $\mathcal{K}(\vec{k}, \A)$. We take $\A$ to be a $d-$ dimensional rectangle parallelepiped, oriented according the Cartesian coordinates, and of length $l_1\times l_2...\times l_d$. Then the kernel $\mathcal{K}$ takes the form:
\begin{equation}
\mathcal{K}(\vec{k}, \A) =\prod\limits_{j=1}^d l_j f_F(k_j, l_j), 
\end{equation}
where $f_F$ is called the Fej\'er kernel. It is a recurring function in interference problems that, as a remarkable example, appeared in Ref. \onlinecite{Wolf2006} in the computation of bounds for the vNEE, that were used to check the violation of the area law for two-dimensional free fermions with a one-dimensional Fermi surface. It has two remarkable convenient properties: first, it has a fairly simple expression in terms of Fourier coefficients:
\begin{align}
f_F(k,l)=\frac{\sin^2(\frac{k l}{2})}{l \sin^2(\frac{k}{2})}=\sum\limits_{r=-l}^l (1-\frac{|r|}{l})e^{i (r k)}.\label{eq:FejFou}
\end{align}
Secondly, it is a uniform approximation of the Dirac delta for convolutions. For such a subsystem $\A$, the bipartite fluctuations can be expressed as:
\begin{equation}
\F_{\alpha \alpha}=\frac{q_e V_{\A}}{4} + \frac{q_e}{4}\sum\limits_{r_1=-l_1}^{l_1}...\sum\limits_{r_d=-l_d}^{l_d}||\hat{\textbf{n}}(\vec{r})||^2_{\alpha} \prod_{j=1}^d (l_j-|r_j|)  \label{eq:FlucBasic}
\end{equation}
This expression naturally expands into $d+1$ different sums, that takes the following schematic form:
\begin{equation}
Q_{d-m}(\vec{l}) \sum\limits_{r_1=-l_1}^{l_1}...\sum\limits_{r_d=-l_d}^{l_d} P_m(\vec{r})||\hat{\textbf{n}}(\vec{r})||^2_{\alpha} \text{ for $0\leq m \leq d$.}
\end{equation}
$P_m$ ($Q_m$) are polynomials consisting in a sum of monomials of degree $m$. Determining the scaling laws of the fluctuations relies on evaluating these sums and the convergence speed of $||\hat{\textbf{n}}(\vec{r})||^2_{\alpha}$. The links between classification of the scaling laws of the bipartite fluctuations and the classification in terms of Sobolev spaces is therefore straightforward.

\subsection{An example: scaling laws in one dimension}\label{App:Ex1D}
We here summarize for reference, and as an introductory example for one dimensional systems, some results that can be found in Ref. \onlinecite{herviou3}.\\
Let us consider the example of the Kitaev chain\citep{Kitaev2001}. It consists in a wire of spinless fermions, with superconducting $p$-wave pairing induced by proximity effect. It verifies:
$$ \vec{n}(k) = (0, 2\Delta \sin k, -\mu - 2t \cos k), \qquad q_e=1.$$
$\Delta$, taken to be real, is the pairing term, $t>0$ describe hopping between neighbouring sites, and $\mu$ is the chemical potential. \\
For $\Delta \neq 0$ and $|\mu|<2t$, the wire is in a topological gapped phase, with winding number $\pm 1$ (it falls in the BDI class), and presents one zero-energy Majorana fermions at each extremity in an open geometry. For $|\mu|>2t$, it is in a trivial gapped phase. There are two families of critical lines, which we are interested in. In the rest, we assume $\mu<0$ for simplicity and compute the fluctuations on a segment of length $l_\A$.\\
The line $\Delta \neq 0$ and $\mu=-2t$ corresponds to a critical $c=\frac{1}{2}$ model in the Ising universality class (one free Majorana mode). $\tilde{n}(k)$ is discontinuous only in $k=0$: $\tilde{n}(0^+)=\vec{e}_y=-\tilde{n}(0^-)$. As a consequence, we obtain:
$$ C^2_{ZZ}(r)=\frac{1}{4\pi^2 r^2} + ... \text{ and } \F_{ZZ}(\A) = i_{ZZ} l_\A - \frac{1}{2\pi^2} \ln l_\A.$$
On the other hand, $\Delta=0$ and $|\mu|<2t$ is a line of free fermions with
$$\tilde{n}(k) = \text{sgn}(-\mu-2t \cos k) \vec{e}_z.$$ It is discontinuous at $\pm k_F$, with $k_F=\arccos(-\mu/2t)$ the Fermi momentum. As an immediate consequence, $$ C^2_{ZZ}(r)=-\frac{\cos^2 2k_F r}{\pi^2 r^2} + ... \text{ and } \F_{ZZ}(\A) = \frac{1}{\pi^2} \ln l_\A.$$
Note the ratio of $2$ in the logarithmic coefficient between the two critical line, corresponding to the ratio of central charges for these conformal models.

\section{QFID} \label{app:QFID}
We define a state $\Ket{\Psi}$ to be $r$-producible in momentum space if:
\begin{equation}
\Ket{\Psi}=\bigotimes\limits_{m=1}^{N/r} \Ket{\psi_m}, \text{ with } \Ket{\psi_m} = f(c^\dagger_{k_{m, 1}}, ...,  c^\dagger_{k_{m, r}}) \Ket{0},
\end{equation}
or in other words, if $\Ket{\Psi}$ is the tensor product of states involving $r$ fermions. Note that here $N$ is the total number of fermionic operator (and not sites). Then, for any observable that can be written $\hat{O}= \sum\limits_{k \in \BZ} \hat{O}_{k}$, one has the bound:
\begin{equation}
\Braket{\hat{O}^2}_c = \sum\limits_{m=1}^{N/r}\Braket{ (\sum\limits_{j=1}^r \hat{O}_{k_{m_j}})^2} \leq \frac{L}{r} \times \frac{r^2}{4}(O_{\text{max}}-O_{\text{min}})^2,
\end{equation}
where $O_{\text{max}/\text{min}}$ is the largest/smallest eigenvalue of $\hat{O}_{k}$. The definition in real space is identical up to the basis change. One can apply this bound for our superconductors and insulators. We limit ourselves to charge and pseudo-spin density fluctuations, but the bounds will be valid for any polarization. For superconductors, $N$ is the actual number of sites, while $\hat{O}=\hat{Q}= \sum\limits_k c^\dagger_k c_k$ such that  $O_{\text{max}}-O_{\text{min}}=1$, leading to:
\begin{equation}
i_{\hat{Q}} \leq \frac{r}{4}.
\end{equation}
For insulators, we need to slightly adapt our conventions to take into account the two fermions by unit-cell properly, and obtain the bound
\begin{equation}
i_{\hat{O}} \leq \frac{r}{2}.
\end{equation}
The two-band non-interacting systems such as the ones studied in this paper are always $2$-producible, which lead to the universal bound
\begin{equation}
i_{\hat{O}} \leq \frac{q_e}{2}. \label{eq:II-BoundQFID}
\end{equation}
Additionally, if $i_{\hat{O}}>\frac{q_e}{4}$, the linear term proves that the ground state is not $1$-producible in real or momentum space, that is to say a simple tensor product of one-fermion wave functions. 

\section{Technical details on computations of $\bc_{\alpha \alpha}$}\label{appen_technical}

\subsection{General considerations}
\subsubsection{Representation in terms of entangling surfaces}
Both for comparison, and for simplicity in computing the contribution of a single corner, it can be convenient to reexpress the bipartite fluctuations in terms of entangling surfaces. Let us consider Eq.~\eqref{fluctuations_form} for bipartite charge fluctuations taken in the continuum limit $\sum\limits_{\vec{r}_1, \vec{r}_2 \in \A}  \to \int\limits_{\A} d^d\vec{r}_1 \int\limits_{\A} d^d\vec{r}_2$, so that
\begin{equation}
  \F_{\alpha \alpha}(\A)= \int\limits_{\A} d^d\vec{r}_1 \int\limits_{\A} d^d\vec{r}_2 f(|\vec{r}_1-\vec{r}_2|)
\end{equation}
when $\C^2_{\alpha \alpha}(\vec{r}) = f(|\vec{r}|)$ is an isotropic function. By applying twice the divergence theorem, we obtain
\begin{widetext}
\begin{equation}
\int\limits_{\A} d^d\vec{r}_1 \int\limits_{\A} d^d\vec{r}_2 f(|\vec{r}_1-\vec{r}_2|) = - \int\limits_{\partial \A} ds_1 \int\limits_{\partial \A} ds_2 \frac{\vec{n}_1.(\vec{r}_1-\vec{r}_2)}{||\vec{r}_1-\vec{r}_2||}\frac{\vec{n}_2.(\vec{r}_1-\vec{r}_2)}{||\vec{r}_1-\vec{r}_2||} F(|\vec{r}_1-\vec{r}_2|) \text{ with } F(r) = \frac{1}{r^{d-1}} \int\limits_0^r dr_1 \int\limits^{r_1}_0 dr_2 r_2^{d-1} f(r_2) \label{eq:App-Entangling}
\end{equation}
\end{widetext}
where $\partial \A$  denotes the boundary of  $\A$. Interestingly, this expression integrates only over the boundaries of $\A$ so that we can isolate the contributions to the logarithmic scaling of the different corner angles and compute them separately.

\subsection{Two dimensions}

\subsubsection{Extracting corner contributions}\label{app:corner}
The contribution of a single corner is derived starting from Eq. \ref{eq:App-Entangling} and keeping only the boundaries $\partial \A_1$ and $\partial \A_2$, making an angle $\psi$, represented in Figure \ref{fig2}. Taking $f(r) = \frac{q_e m^2}{16 \pi^2} r^{-4}+...$ corresponds to $F(r) = \frac{q_e m^2}{32 \pi^2} r^{-2}+...$,
or $F(r) = \frac{q_e m^2}{8 \pi^2} \frac{r^2}{(1+r^2)^2}$ which gives the same logarithmic term. We have now to evaluate
\begin{multline}
2\int\limits_0^{l} \int\limits_0^{l}  dx_1 dx_2 \frac{x_1 x_2 \sin^2 \alpha}{x_1^2+x_2^2-2 x_1 x_2 \cos \alpha}\\ F(\sqrt{x_1^2+x_2^2-2 x_1 x_2 \cos \alpha})
\end{multline}
Taking the derivative with respect to $l$ and computing the integral at large $l$ gives
\begin{equation}
\frac{q_e m^2}{32 \pi^2}\times (1+(\pi-\alpha)\cot \alpha)\times \frac{1}{l}
\end{equation}
corresponding to the corner function Eq. \ref{exactCorner} in the main text.

\subsubsection{Computation for a parallelogram}\label{app:Para}
The starting point is given in Eq. \ref{eq:seriespara} that we reproduce here:
\begin{equation}
\mathfrak{S}(l_u, l_v, \vec{u}, \vec{v})= \sum\limits_{r_v=-l_v}^{l_v}  |r_u||r_v| ||\hat{\textbf{n}}(r_u \vec{u} + r_v \vec{v})||_\alpha^2.
 \end{equation}
$$\vec{u}= |u|( \cos \theta_u , \sin \theta_u  ), \vec{v}= |v|( \cos (\theta_u + \psi) , \sin (\theta_u + \psi) )$$

We start by taking the continuum limit, and proceed to a change of variable. We define $(x, y) = r_u \vec{u} + r_v \vec{v} = J (r_u, r_v)$. As $(\vec{u}, \vec{v})$ generates a parallelogram of area $1$, $|\det J|=1$, and one obtains:
\begin{multline}
\mathfrak{S}(l_u, l_v, \vec{u}, \vec{v}) = \iint\limits_{P}\frac{r^3}{\sin \psi} |\sin(\theta-\theta_u-\psi)\sin(\theta-\theta_u)|\\
||\hat{\textbf{n}}(\vec{r})||_\alpha^2 dr d\theta,
\end{multline}
where the integrals carry on the parallelogram $P$ centered in $\vec{0}$, of side $2l_u \vec{u}$ and $2l_v \vec{v}$. As only the leading term in $||\hat{\textbf{n}}(\vec{r})||_\alpha^2$ gives the logarithmic contribution, one can replace it by the test function:
\begin{equation}
g_{n, \varepsilon}(\vec{r}) = \frac{\cos (n \theta)}{(r^2+e^2)^2}.
\end{equation}
As $g_{n, \varepsilon}(\vec{r}) ||\hat{\textbf{n}}(\vec{r})||_\alpha^{-2} \rightarrow C + O(r^{-1})$, with $C \neq 0$, the logarithmic coefficient induced by $||\hat{\textbf{n}}(\vec{r})||_\alpha^2$ is simply $C$ times the one generated by $g_{n, \varepsilon}(\vec{r})$.
Integration on a parallelogram is tricky in this naturally polar expression, so we introduce
\begin{align*}
R_{\text{max}}^2&=\text{max } l_u^2+l_v^2\pm 2 l_u l_v u v \cos \psi \\
R_{\text{min}}&=\text{min } (l_u u \sin \psi, l_v v \sin \psi)
\end{align*}
The disc $D$ of radius $R_{\text{min}}$ ($R_{\text{max}}$) is inscribed in $P$ (circumscribes it), and one obtains:
\begin{multline*}
|\iint\limits_{P\setminus D(R_{\text{min}})} \frac{r^2}{\sin \psi} |\sin(\theta-\theta_u-\psi)\sin(\theta-\theta_u)| g_{n, \varepsilon}(\vec{r})d\vec{r}|\\
\leq \frac{4l_u l_v - \pi R_{\text{min}}^2 }{ R_{\text{min}}^2 \sin \psi} 
\end{multline*}
\begin{multline*}
|\iint\limits_{D(R_{\text{max}}) \setminus P} \frac{r^2}{\sin \psi} |\sin(\theta-\theta_u-\psi)\sin(\theta-\theta_u)| g_{n, \varepsilon}(\vec{r})d\vec{r}|\\
\leq \frac{\pi R_{\text{max}}^2-4l_u l_v  }{R_{\text{min}}^2 \sin \psi}
\end{multline*}
Finally, a simple computation leads to:
\begin{widetext}
\begin{equation*}
\iint\limits_{D(R)}\frac{r^2}{\sin \psi} |\sin(\theta-\theta_u-\psi)\sin(\theta-\theta_u)| g_{n, \varepsilon}(\vec{r})d\vec{r}= \ln R \int\limits_{0}^{2 \pi} \frac{d \theta}{\sin  \psi}  |\sin(\theta-\theta_u-\psi)\sin(\theta-\theta_u)| \cos n \theta + \mathcal{O}(1)
\end{equation*}
\begin{equation}
\int\limits_{0}^{2 \pi} \frac{d \theta}{\sin  \psi}  |\sin(\theta-\theta_u-\psi)\sin(\theta-\theta_u)| \cos n \theta = 
\left\lbrace \begin{array}{c}
0 \text{ for } n \text{ odd} \\
2+(\pi-2\psi) \cot \psi \text{ for } n=0 \\
-\frac{\pi-2 \psi+\sin 2 \psi}{2\sin \psi} \cos(2 \theta_u + \psi) \text{ for } n=\pm 2 \\
\frac{2}{m^2-1} (- \cos (m \psi) + \frac{\sin m \psi}{m} \cot \psi) \cos (m(2 \theta_u + \psi))  \text{ for } n=\pm 2m
\end{array} \right.\nonumber
\end{equation}
\end{widetext}
At fixed $u, v, \psi$, and in the limit where $l_u$ and $l_v$ grow at a similar pace when $\A \rightarrow S$, we directly obtain the coefficients given in the main text.\\

For an asymmetric growth of $\A$, one can prove that the logarithmic term can be replaced by:
\begin{equation}
\frac{1}{2}\ln \frac{l_u^2l_v^2}{l_u^2+l_v^2}
\end{equation}

\subsection{Three dimensions}
\subsubsection{Computation of $\hat{\textbf{n}}(\vec{r})$} \label{app:WPCorr}
We propose two possible computations of the Fourier transform of the flattened Hamiltonian $\tilde{\textbf{n}}(\vec{k})$ for a three-dimensional Weyl point. The first one is a direct computation using rotational invariance. The second method takes advantage of the plane wave expansion in terms of spherical harmonics. We assume an underlying cubic lattice for simplicity.\\

Let us first consider a single isotropic WP of Chern number $\pm 1$ given in Eq. \ref{eq:HamWP}. The singular contribution of the flattened Hamiltonian $\tilde{\textbf{n}}(\vec{k})$ is well captured by the regularized function:
\begin{equation}
g(\vec{k}) =\left\lbrace \begin{array}{c}
D_{\pm 1}\frac{\vec{k}}{k}(1-k^2)^m, \text{ with } m\geq 2, \text{for } k\leq 1 \\
0 \text{ else}
\end{array} \right.
\end{equation}
To compute its Fourier transform, one can take advantage of the rotational invariance. Let $R$ a rotation that send $\vec{r} = r R \vec{e}_z$. Then, one  obtains
\begin{equation}  \label{eq_gr}
  \begin{split}
\hat{g}(\vec{r}) &= D_{\pm 1} R\iiint\limits_{\BZ} \frac{\vec{k}}{k}(1-k^2)^m e^{ikr\cos \theta} \frac{d\vec{k}}{8 \pi^3}\\
&= D_{\pm 1} \frac{\vec{r}}{4 \pi^2 r}\int\limits_0^1 dk \int\limits_0^\pi d\theta \sin \theta \cos \theta (1-k^2)^m e^{ikr\cos \theta}\\
&\approx  D_{\pm 1} \frac{i \vec{r}}{\pi^2 r^4} + \mathcal{O}(r^{-4})
  \end{split}
\end{equation}
corresponding to the large distance asymptote for $\hat{\textbf{n}}(\vec{r})$ given in Eq.~\eqref{eq_nr}.

Alternatively, we note that the flat-band Hamiltonian can be written close to the WP as
\begin{equation}\label{eq_tilden}
  \tilde{n}_{\pm 1}(\vec{k})= \frac{2\sqrt{\pi}} {\sqrt{3}}
\begin{pmatrix} \mp \sqrt{2} \text{Re}(Y_1^{\pm 1}) \\ \mp \sqrt{2\pi} \text{Im}(Y_1^{\pm 1}) \\
Y_1^0  
\end{pmatrix},
\end{equation}
where the $Y_l^m$ denote spherical harmonics.
%The higher-order harmonics appear when considering WP with chirality larger than $1$.
We make use of the expansion
\begin{equation}
e^{i \vec{k}.\vec{r}} = 4 \pi \sum\limits_{l, m}i^lj_l(kr)Y_{l}^m(\vec{r})Y_{l}^{m *}(\vec{k}),
\end{equation}
substituted in the Fourier transform of $\tilde{n}_{\pm 1}(\vec{k})$ (regularized as $g(\vec{k})$) in Eq.~\eqref{eq_gr}, together with the orthogonality identity
\begin{equation}
\iint\limits_{\mathbb{S}^2} Y_{l}^{m *}(\vec{k})Y_{l'}^{m'}(\vec{k}) d \vec{k} = \delta_{l, l'} \delta_{m, m'},
\end{equation}
and the integral
\begin{equation}
\int\limits_{0}^1 j_l(kr) k^2 (1-k^2)^m dk= \frac{2 \sqrt{\pi} \Gamma(\frac{3+l}{2})}{\Gamma(\frac{l}{2})}\frac{1}{r^3}+ \mathcal{O}(r^{-5}),
\end{equation}
to derive the formula
\begin{equation}\label{eq_spherical}
  \begin{split}
    \hat{Y}_l^m (\vec{r}) & = \int\limits \frac{d\vec{k}}{8\pi^3} e^{i\vec{k}.\vec{r}}(1-k^2)^2 Y_l^m(\vec{k}) \\
    & =\frac{\Gamma(\frac{3+l}{2})}{\pi \sqrt{\pi} \Gamma(\frac{l}{2})}  \frac{i^l}{r^3}Y_l^m(\vec{r}),
  \end{split}
\end{equation}
recovering the result of Eq.~\eqref{eq_gr} for $l=1$ and $m=-1,0,1$.

In the case of higher chiralities $|m| >1$, the flattened Hamiltonian has a form similar to Eq.~\eqref{eq_tilden} but involving spherical harmonics of higher degrees. Nevertheless, the above second method can be also used for these chiralities with the result Eq.~\eqref{eq_spherical}.

\subsubsection{Logarithmic contribution} \label{app:WPLog}
In this Section, we extract the logarithmic contribution to the fluctuations in a three dimensional materials, induced by $r^3$ scaling terms. \\
Let us first consider $\A$ to be a sphere of radius $R$. It is convenient to first study the spin-spin (Hei) fluctuations, whose long-range behavior is captured by the test function:
\begin{equation}
g_\varepsilon (\vec{r})=\frac{1}{(r^2 + \varepsilon^2)^3}
\end{equation}
and the fluctuations can be obtained by computing:
\begin{align}
C &= \int\limits_{\A^2} d\vec{r}d\vec{r}' g_\varepsilon(\vec{r}-\vec{r}')\\
&= 2\pi^2 \int\limits_0^R dr \int\limits_0^R dr'(\frac{r r'}{(r-r')^2+\varepsilon^2)^2}-\frac{r r'}{(r+r')^2+\varepsilon^2)^2}) \nonumber\\
&=\frac{\pi^2}{6} ( \frac{R^3}{\varepsilon^3}( \pi - 2 \arctan(\frac{\varepsilon}{2R})+2 \arctan(\frac{2R}{\varepsilon})) \nonumber \\
 &\qquad -4 \frac{R^2}{\varepsilon^2}+ \ln (1+4 \frac{R^2}{\varepsilon^2})),
\end{align}
which leads to the coefficient given in the main text.

\section{Relation between entropy and fluctuations for Dirac fermions}\label{app:comparison}
In this Section, we provide some supplementary details on the (lack of) relation between entropy and  charge fluctuations for two-dimensional Dirac fermions.\\
While most of the considered charges are not conserved, a direct comparison between the logarithmic contributions to the entropy and the fluctuations is still in order. As the corner functions differ, it is a priori impossible to obtain a constant ratio as in Eq. \ref{eq:RatioFermiGas}. There are, though, several limits one could consider (namely $\psi\rightarrow 0, \frac{\pi}{2}, \pi$). Using the values given in Refs. \onlinecite{Bueno2015-3, Helmes2016}, we obtain Table \ref{tab:Ratio}. No simple relations can be extracted from these ratios.
\begin{table}
\begin{tabular}{|c|c|c|c|c|} \hline
$\psi$ & $\alpha=1$ & $\alpha=2$ & $\alpha=3$ & $\alpha=4$ \\ \hline
$0$ & $ 7.26 $ & $ 4.75 $ & $ 4.09 $ & $ 3.79 $\\ \hline
$\frac{\pi}{2}$ & $ 7.36 $ & $ 4.73$ & $4.05 $ & $3.74$\\ \hline
$\pi$ & $ \frac{3}{4} \pi^2$ & $\frac{3}{2} \pi $ & $ \frac{20}{9\sqrt{3}}\pi $ & $\frac{1+6\sqrt{2}}{8} \pi $\\ \hline
\end{tabular}
\caption{Ratio between logarithmic contributions in the von Neumann ($\alpha=1$) and the first Renyi entropies $\Ss_\alpha = \frac{1}{1-\alpha} \ln \text{Tr} \rho^\alpha$ and the bipartite fluctuations $-\frac{a_{\Ss_{\alpha}}(\psi)}{a_{ZZ}(\psi)}$ for a single Dirac cone with winding number $\pm 1$, for the three angles $0$, $\frac{\pi}{2}$ and $\pi$. No simple relation of the form given in Eq. \ref{eq:RatioFermiGas} emerges at any of these angles.}
\label{tab:Ratio}
\end{table}

\bibliography{PapierFluctuations3}

\end{document}